\documentclass[aps,
prx,
amsmath,
amssymb,
floatfix,
superscriptaddress,
reprint,
draft,
final,
]{revtex4-2}
\usepackage[utf8]{inputenc}
\usepackage[OT1]{fontenc}
\usepackage{graphicx}   
\usepackage{xpunctuate}
\usepackage{siunitx}
\usepackage{physics}
\usepackage[super]{nth}
\usepackage{color}      
\usepackage{soul}
\usepackage{hyperref}   
\usepackage[capitalize]{cleveref}

\def\clap#1{\hbox to 0pt{\hss#1\hss}}

\def\mathrlap{\mathpalette\mathrlapinternal}

\def\mathrlapinternal#1#2{%
	\rlap{$\mathsurround=0pt#1{#2}$}}

\newcommand*{\hb}{\hbar}
\newcommand*{\oh}[1]{\hat{#1}}

\begin{document}

\title{Simulating long-range hopping with periodically-driven superconducting qubits}

\author{Mor M. \surname{Roses}}
\affiliation{Department of Physics and Center for Quantum Entanglement Science and Technology, Bar-Ilan University, 52900 Ramat-Gan, Israel}

\author{Haggai \surname{Landa}}
\affiliation{IBM Quantum, IBM Research Haifa, Haifa University Campus, Mount Carmel, Haifa 31905, Israel}

\author{Emanuele G. \surname{Dalla Torre}}
\affiliation{Department of Physics and Center for Quantum Entanglement Science and Technology, Bar-Ilan University, 52900 Ramat-Gan, Israel}

\begin{abstract}
Quantum computers are a leading platform for the simulation of many-body physics. This task has been recently facilitated by the possibility to program directly the time-dependent pulses sent to the computer. Here, we use this feature to simulate quantum lattice models with long-range hopping. Our approach is based on an exact mapping between periodically driven quantum systems and one-dimensional lattices in the synthetic Floquet direction. By engineering a periodic drive with a power-law spectrum, we simulate a lattice with long-range hopping, whose decay exponent is freely tunable.
We propose and realize experimentally two protocols to probe the long tails of the Floquet eigenfunctions and identify a scaling transition between  {long-range and short-range} couplings. Our work offers a useful benchmark of pulse engineering and opens the route towards quantum simulations of rich nonequilibrium effects.
\end{abstract}

\maketitle

A common assumption of many-body physics is that particles can interact only with their neighbors. Similarly, in canonical lattice models, particles are often assumed to hop only between neighboring sites. While generically correct in condensed matter physics, this assumption is not applicable to gravitational models, unscreened Coulomb interactions, and synthetic materials, where the couplings can decay as a power-law. Long-range couplings can lead to counter-intuitive effects, such as violations of thermodynamic identities \cite{morigi2004eigenmodes,bouchet2010thermodynamics,gupta2017world,campa2009statistical,campa2014physics} and anomalous mechanisms of information spreading \cite{richerme2014non,zhang2017observation,vzunkovivc2018dynamical}. For systems with power-law decaying couplings, the transition between long range and short range correlations occurs at $\alpha=d$, where $\alpha$ is the decay power of the potential and $d$ is the dimension \cite{dyson1969existence,thouless1969long,sak1973recursion}. This transition is further affected by disorder \cite{levitov1999critical,nag2019many}, thermal \cite{brezin2014crossover,behan2017long} and quantum \cite{defenu2020criticality} fluctuations, and competing orders \cite{campa2019ising,xiao2019competition}, giving rise to a wide range of classical and quantum phases with peculiar scaling laws.

The experimental study of this transition requires a simulator where $\alpha$ is not set by the physical decay of elementary forces and can be tuned continuously. This requirement is partially fulfilled by trapped ions, where phonon-mediated interactions can be used to simulate long-range quantum spin models, and $\alpha$ can be tuned within a limited range \cite{porras2004effective,islam2013emergence}. Here, we propose and realize an alternative approach, based on periodically driven (Floquet) quantum models. In these systems, the frequency components of the wavefunctions give rise to an effective one-dimensional lattice and the harmonics of the drive correspond to arbitrary hopping terms \footnote{see Ref.~\cite{rudner2020floquet} for an introduction}. Intuitively, this synthetic Floquet dimension describes the number of photons absorbed or emitted from the drive. In the theoretical limit of a classical drive, such dimension is infinite in both directions.

To realize our protocol, we use a feature of noisy intermediate-scale quantum (NISQ) computers that has been recently made available via the cloud, namely \emph{pulse engineering}. This term refers to the possibility of feeding arbitrary time-dependent pulses to the device \cite{cross2017open,mckay2018qiskit,aleksandrowicz2019qiskit}.
Pulse engineering is commonly used to characterize the qubits and to optimize the fidelity of a target gate, or of a set of gates  \cite{khaneja2005optimal,de2011second,kelly2014optimal,glaser2015training,lu2017enhancing,gokhale2019partial,chen2014fidelity,bukov2018reinforcement,zhang2019does,niu2019universal,stenger2020simulating,alexander2020qiskit,garion2020experimental}. Periodic drives have been recently used to create long-range couplings in the physical space \cite{bastidas2020fullyprogrammable} and to simulate a two dimensional topological insulator \cite{malz2020topological}.  In this work, we use pulse engineering to simulate long-range couplings in the Floquet space.

Floquet theorem \cite{floquet1883equations} guarantees that the time evolution of a system governed by a time-periodic Hamiltonian, $\oh{H}(t+\tau)=\oh{H}(t)$,  can be written as $\ket{\psi(t)} = \sum_n c_n e^{-i\mu_n t}\ket{\phi_n(t)}$. Here, $\mu_n$ are the quasienergies, and  $\ket{\phi_n(t)}$ are the Floquet functions. These functions are periodic in time, $\ket{\phi_n(t+\tau)}=\ket{\phi_n(t)}$, and can be expanded in a discrete Fourier series,
\begin{align}
\label{eq:phi_dft}
\ket{\phi_n(t)}=&\sum_m e^{-i m\omega_pt}\ket{\phi_n(m)},
\end{align}
where $\omega_p=2\pi/\tau$ is the pump frequency. By integrating the Schr\"odinger equation over one period, one obtains a \emph{time independent} linear system of equations $\oh{H}_F \ket{\phi_n} = \hb\mu_n \ket{\phi_n}$, where $\oh{H}_F$ is the Hamiltonian in an enlarged infinite dimensional Floquet space and $\ket{\phi_n}$ is a vector in this space, with $\braket{m}{\phi_n}=\ket{\phi_n(m)}$.
The central block of $\oh{H}_F$ is given by
\begin{align}
        \spmqty{
                \oh{H}_0 + 2\hb\omega_p && \oh{H}^{(1)}        && \oh{H}^{(2)}  && \oh{H}^{(3)}        && \oh{H}^{(4)}          \\\\
                \oh{H}^{(-1)}        && \oh{H}_0 + \hb\omega_p && \oh{H}^{(1)}  && \oh{H}^{(2)}        && \oh{H}^{(3)}          \\\\
                \oh{H}^{(-2)}        && \oh{H}^{(-1)}       && \oh{H}_0      && \oh{H}^{(1)}        && \oh{H}^{(2)}          \\\\
                \oh{H}^{(-3)}        && \oh{H}^{(-2)}       && \oh{H}^{(-1)} && \oh{H}_0 - \hb\omega_p && \oh{H}^{(1)}          \\\\
                \oh{H}^{(-4)}        && \oh{H}^{(-3)}       && \oh{H}^{(-2)} && H^{(-1)}       && \oh{H}_0 - 2 \hb\omega_p \\\\
        },
\label{eq:HF}
\end{align}
where $\oh{H}^{(m)}=\tau^{-1}\int_{0}^\tau \oh{H}(t)e^{i m\omega_pt}\dd{t}$ is the $m$\textsuperscript{th} discrete Fourier component of $\oh{H}(t)$ \cite{shirley1965solution}. In the limit of large $\omega_p$, the off-diagonal terms of $\oh H_F$ can be treated perturbatively, giving rise to the well-known Magnus expansion. Note that \cref{eq:phi_dft} is invariant under the gauge transformation $\ket{\phi_n(m)}\rightarrow e^{i\omega_pt}\ket{\phi(m+1)}$, and hence $\oh{H}_F$ is invariant to discrete translations, $m\to m+1,~\oh{H}_F\to \oh{H}_F+\hb\omega_p$.

Textbook examples of Floquet systems often involve sinusoidal drives, where $\oh{H}^{(m)}$ is non zero only for $m=\pm 1$, and $\oh{H}_F$ corresponds to a tilted lattice with nearest-neighbor couplings. In this case, all the eigenstates of $\oh{H}_F$ are localized in Floquet space. In analogy to Bloch oscillations of electrons in an electric field, quantum particles oscillate in the Floquet dimension, leading to a periodic alternation of energy emission and absorption (Rabi oscillations) \cite{russomanno2017floquet,gagge2018bloch}. The situation does not change qualitatively when a finite number of harmonics is added. Hence, in most physical situations it is sufficient to focus only on a small segment of the Floquet space \footnote{Remarkably, this approximation remains valid even in situations where the system shows a dynamical instability and the energy absorption rate diverges. For example, the parametric resonance of an harmonic oscillator can be described by considering only two Fourier harmonics \cite{landau1976mechanics}. At the resonance, the Floquet functions diverge in Fock space of the oscillator, but remain localized in  Floquet space \cite{weigert2002quantum}.}. The opposite situation is encountered if the system is periodically driven by delta-functions in time, giving rise to celebrated kicked models \cite{chirikov2008chirikov}, where $\oh{H}^{(m)}$ is independent on $m$. The Hamiltonian $\oh{H}_F$ corresponds to a model with all-to-all couplings and its eigenfunctions are completely delocalized in the Floquet space. In this case,
a truncation of the Floquet space can lead to unphysical predictions \footnote{The delocalization of the Floquet eigenfunctions is a necessary condition for the dynamical localization in quantum models of kicked rotors \cite{grempel1982localization,fishman1982chaos}.}.

\begin{figure}[t!]
	\centering
	\includegraphics[width=\linewidth]{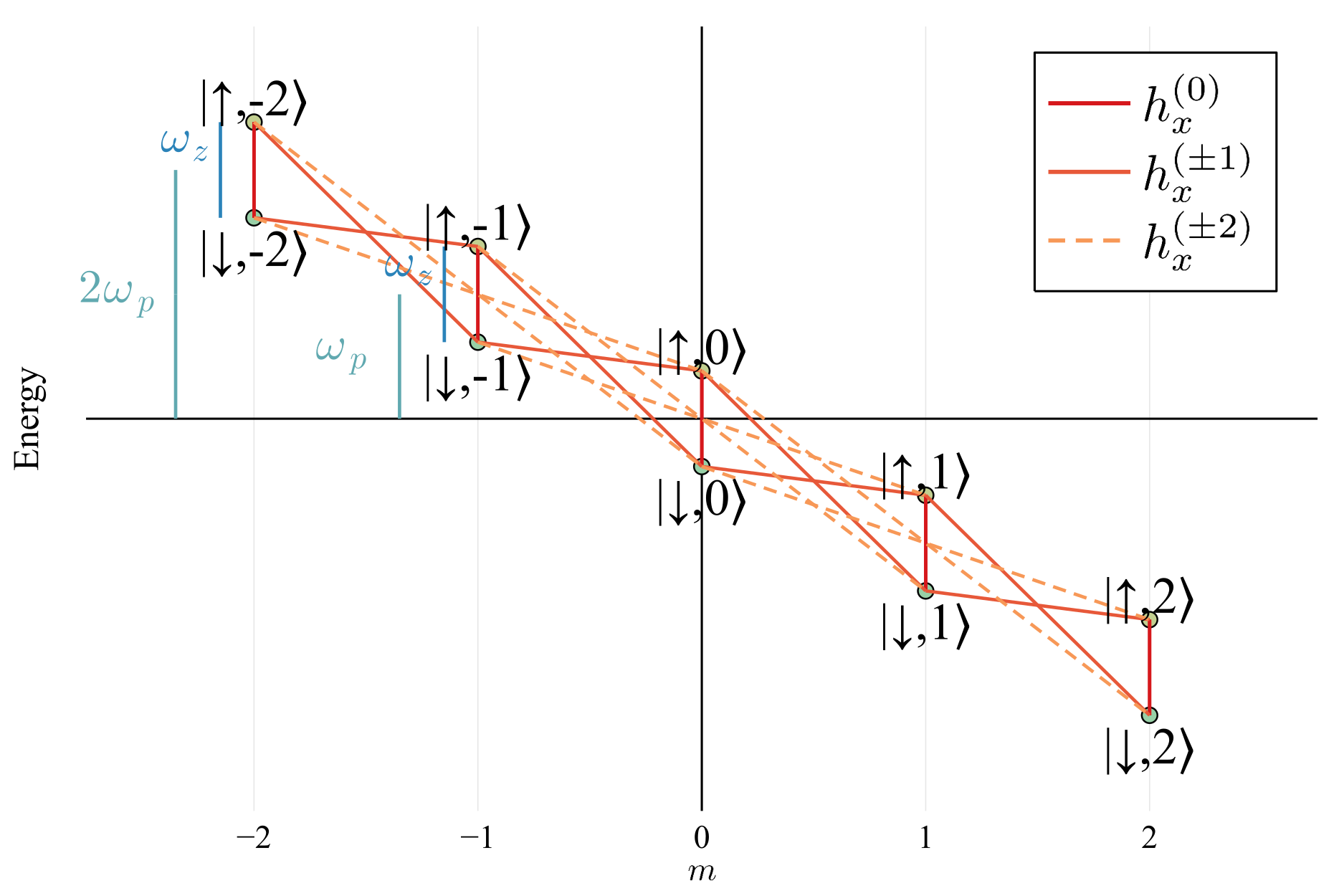}
	\caption{%
		Schematic plot of the Floquet Hamiltonian for a single qubit described by \cref{eq:Ht} with $h_x(t)=h_x(t+\tau)$.
		Each state is characterized by two quantum numbers: spin ($\uparrow$ or $\downarrow$) and Floquet index ($m$).
		The vertical lines show the energy scales: the driving frequency, $\omega_p=2\pi/\tau$, and the two-level splitting, $\omega_z$.
		The first and second harmonics of the drive $h_x^{(\pm1)}$ and $h_x^{(\pm2)}$ correspond, respectively, to nearest neighbour and next-nearest neighbour couplings.
	}
	\label{fig:energy-levels}
\end{figure}

Aiming at quantum simulations with superconducting qubits, we specifically consider the spin model
\begin{align}
        \label{eq:Ht}
        \oh{H}(t) = & -\frac{\hb\omega_z}2\oh{\sigma}_z + \frac{\hb h_x(t)}2\oh{\sigma}_x,
\end{align}
where $\omega_z$ and $h_x(t)$ are tunable {fields}, and $\sigma_\alpha$ are Pauli matrices \footnote{The Hamiltonian (\ref{eq:Ht}) is actually realized in a rotating frame, such that $\omega_z$ is the detuning between the driving field and the bare frequency of the qubit, and $h_x(t)$ is one quadrature of the driving field. We point out that the other quadrature enables one to realize lattice model with imaginary hopping terms.}. For $h_x(t)=h_x(t+\tau)$ one obtains a Floquet Hamiltonian of the form of \cref{eq:HF}, with $\oh{H}_0 = -\hb\omega_z\oh{\sigma}_z/2+\hb h_x^{(0)}\oh\sigma_x/2$ and $\oh H^{(m)}=h_x^{(m)}\oh\sigma^x$, where $h_x^{(m)}=\tau^{-1}\int_0^\tau h_x(t)e^{i m\omega_p t} \dd{t}$. This Hamiltonian is equivalent to a two-band Wannier–Stark ladder and is schematically shown in \cref{fig:energy-levels}, where, for convenience, we have depicted $\oh H^{(m)}$ for $|m|\leq 2$ only. To study long-range hopping models, we introduce a time-dependent drive with a power-law frequency spectrum,
\begin{align}
	h_x^{(m)}=\frac{h_0}{(1+\abs{m})^\alpha}
	\label{eq:floquet_drive}.
\end{align}
This driving field interpolates between a kicked model at $\alpha=0$ and a weak sinusoidal drive for $\alpha\gg 1$.  At intermediate $\alpha$'s, the Fourier transform of \cref{eq:floquet_drive} corresponds to periodic kicks with a finite width. The resulting time evolution can be easily computed in two limiting cases: (i) For $\alpha=0$, the time evolution over one period is $U(\tau)=U_zU_x$, {where $U_z=\exp(-i\omega_z\tau\oh\sigma_z/2)$ corresponds to the evolution between the kicks and $U_x=\exp(-i h_0\tau\oh\sigma_x/2)$ describes the kicks. Note that the same stroboscopic time evolution can be obtained by applying alternatively magnetic fields in the $x$ and $z$ direction for finite times. Such bang-bang\cite{bellman1956bang,viola1998dynamical} protocol and the kicked model are characterized by the same stroboscopic evolution  (which is the evolution over an integer number of periods), but lead to a different micromotion (i.e. the intermediate evolution during the time periods). Due to this difference, these two protocols are mapped to different lattice models: the bang-bang protocol corresponds to a short-range hopping model, while the kicked protocol is mapped to a long-range model.} (ii) For $\omega_z=0$, the Hamiltonian corresponds to a time-dependent magnetic field in the $x$ direction, whose time evolution is $U(t)=\exp(-i\int\dd{t} h_x(t)\oh\sigma_x/2)$.

%
The driving field of \cref{eq:floquet_drive} undergoes a scaling transition between short-range and long-range {as a function of $\alpha$. In general, a system is said to be short ranged (or long-ranged) if the integral over space of the coupling is finite (or infinite). In systems with short-range couplings the total energy is extensive (i.e. proportional to the size of the system), while for long-ranged couplings, the total energy grows faster than the system size. See, for example, Refs. \cite{campa2019ising,defenu2020criticality} for the critical properties of this transition, in the context of particles with power-law interactions. In our mapping to a one-dimensional lattice, the Fourier component $h_x^{(m)}$ plays the role of the coupling between two sites at distance $m$, see \cref{fig:energy-levels}. The system is short/long ranged depending on the finiteness of the sum of $h_x^{(m)}$ over $m$, $h_{\mathrm{sum}}=\sum_m h_x^{(m)}$. In the case of the power-law spectrum introduced in \cref{eq:floquet_drive}, this quantity diverges for $\alpha<1$. Hence, the system under consideration undergoes a transition between short range and long range at $\alpha=1$.

\begin{figure}[t!]
	\centering
	\includegraphics[width=\linewidth]{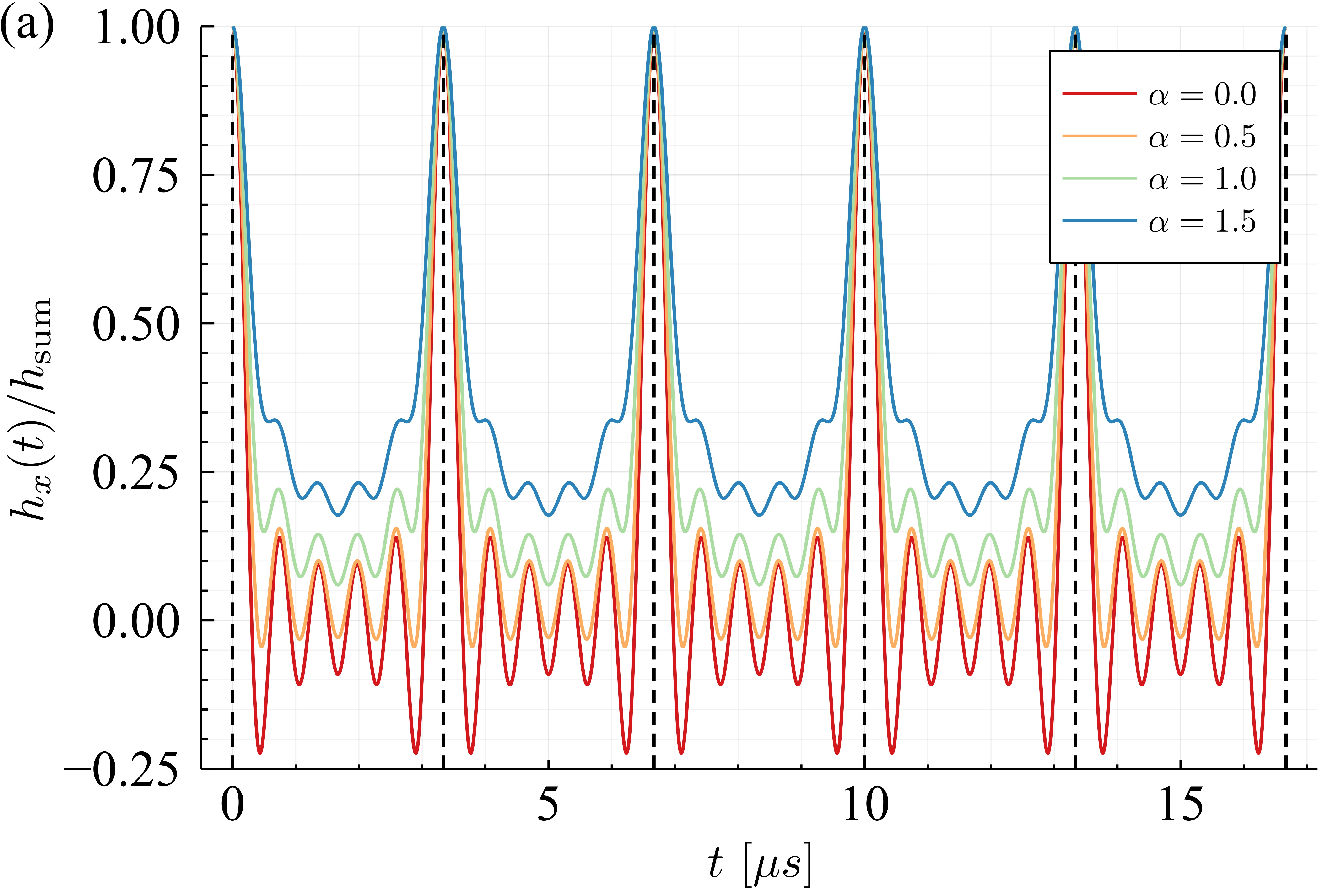}
	\includegraphics[width=\linewidth]{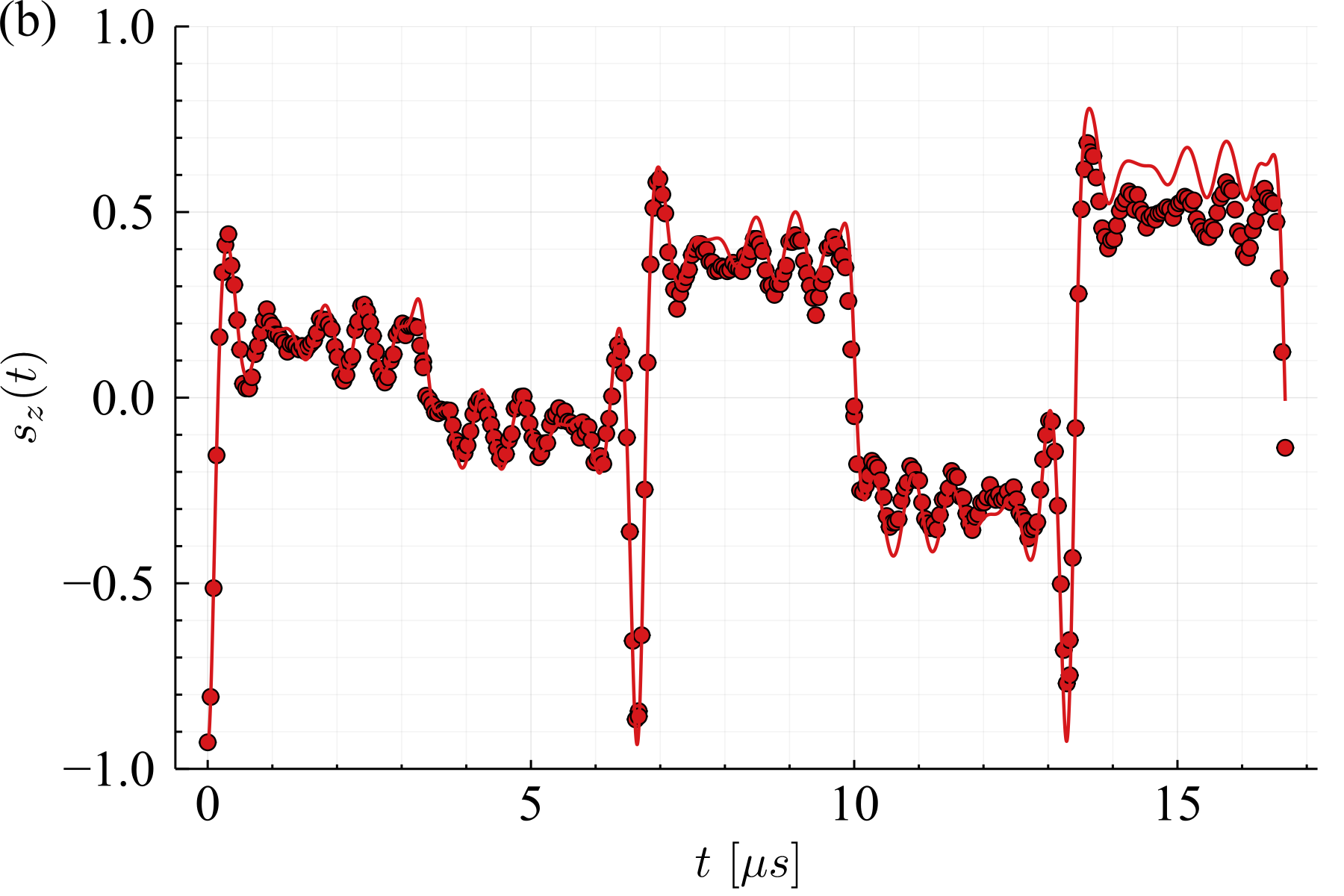}
	\caption[]{%
		(a) Time-dependent drive with power law spectrum, \cref{eq:floquet_drive}, with a cutoff at $M=5$, for different values of $\alpha$. (b) Time evolution of the qubit for $M=5$ and $\alpha=0$, theory (lines) and experiment (dots).
	}
	\label{fig:drive-plot}
\end{figure}

In our mapping of periodically driven dynamics to long-range hopping, $h_\mathrm{sum}$ corresponds to the effective drive strength
at integer multiples of $\tau$, $h_\mathrm{sum}=h_x(n\tau)$ with integer $n$, which is always finite in a real experiment. To address this problem, we} introduce a physical cutoff $M$, by assuming that $h_x^{(m)}$ is given by \cref{eq:floquet_drive} for all $m\leq M$, and equals 0 for $m>M$.
The resulting time dependent drive in shown in \cref{fig:drive-plot}(a).
As we will see, the cutoff parameter $M$ offers a powerful knob to probe the power-law tails of the Floquet functions.

\begin{figure}[t!]
	\centering
	\includegraphics[width=\linewidth]{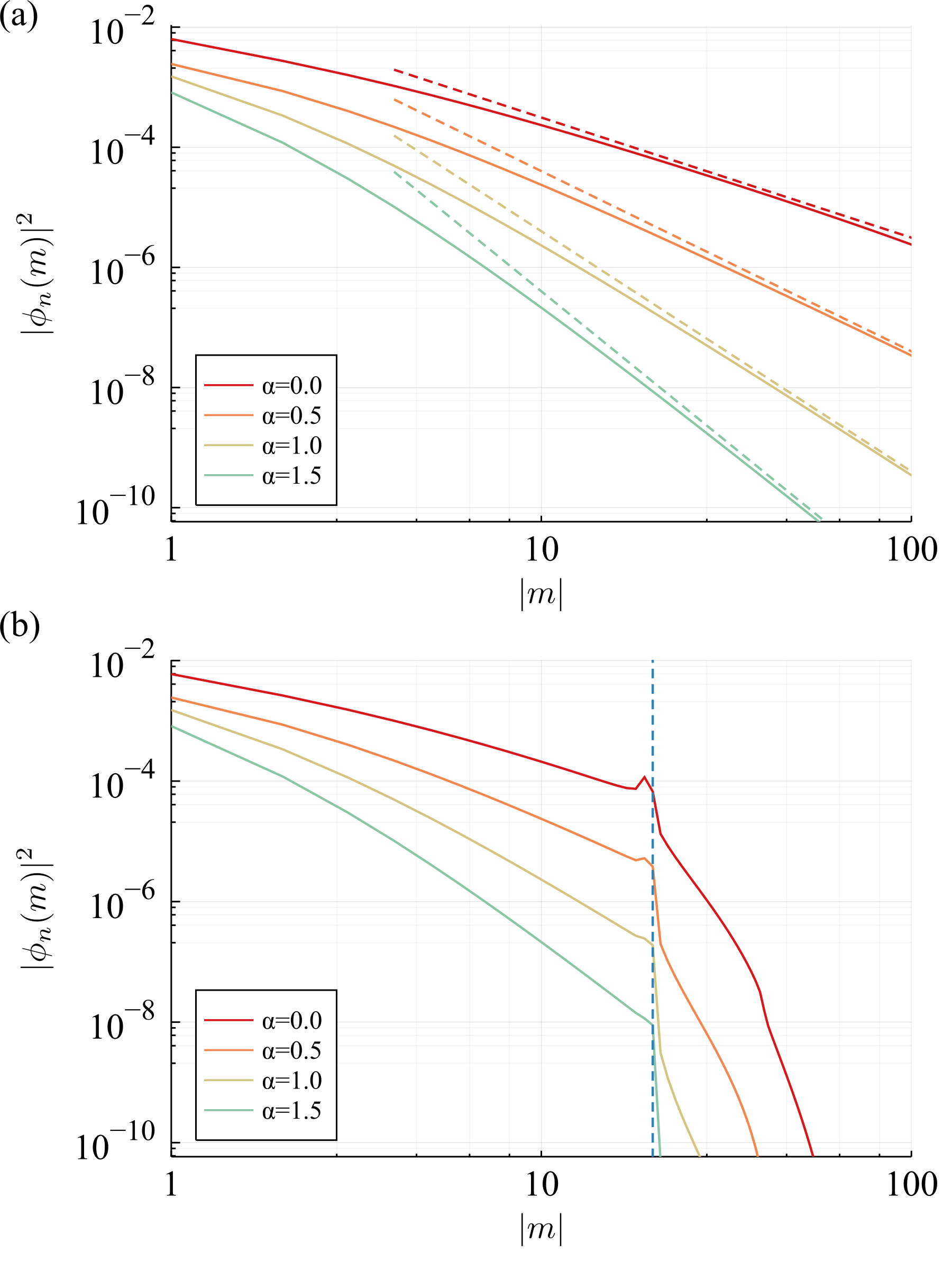}
	\caption[]{%
		(a) Eigenstate of the Floquet Hamiltonian: numerical diagonalization of $\oh H_F$ (continuous lines) and power-law asymptotics (dashed lines), \cref{eq:powerlaw}. The normalization of the wavefunction is guaranteed by the value of $\abs{\phi_n(m=0)}^2$ (not shown). (b) Numerical diagonalization in the presence of a cutoff at $M=20$, giving rise to distinctive peak at $m=M$.}
	\label{fig:alpha}
\end{figure}

Our experimental setup is a single-qubit quantum computer, available on the cloud, the IBM Quantum processor\cite{ibm2021} \verb|ibmq_armonk v2.4.0| which is one of the IBM Quantum Canary processors. The typical physical parameters that we use are $\omega_z=2\pi\times\SI{250}{\kilo\hertz}$, $h_{\mathrm{sum}}=2\pi\times\SI{1.2}{\mega\hertz}$, and $\omega_p=2\pi\times\SI{300}{\kilo\hertz}$, or $\tau=\SI{3.3}{\micro\second}$. This choice enables us to have $\delta t\ll \tau \ll \min[T_1,T_2]$, where $\delta t = \SI{0.00022}{\micro\second}$ is the smallest programmable time step and $T_1=\SI{160}{\micro\second}$ and $T_2=\SI{280}{\micro\second}$ are the decay and coherence times. In our experiment, we prepare the qubit in the $\ket{\downarrow}$ state, set $\omega_z$ to a fixed value, apply the drive $h_x(t)$ for time $t$, measure the expectation value $s_z(t)=\ev*{\oh\sigma_z}$ over 8192 shots, and repeat the procedure for 740 time steps between $t=0$ and $T=5\tau$, see \cref{fig:drive-plot}(b).

\begin{figure}[t!]
	\centering
	\includegraphics[width=\linewidth]{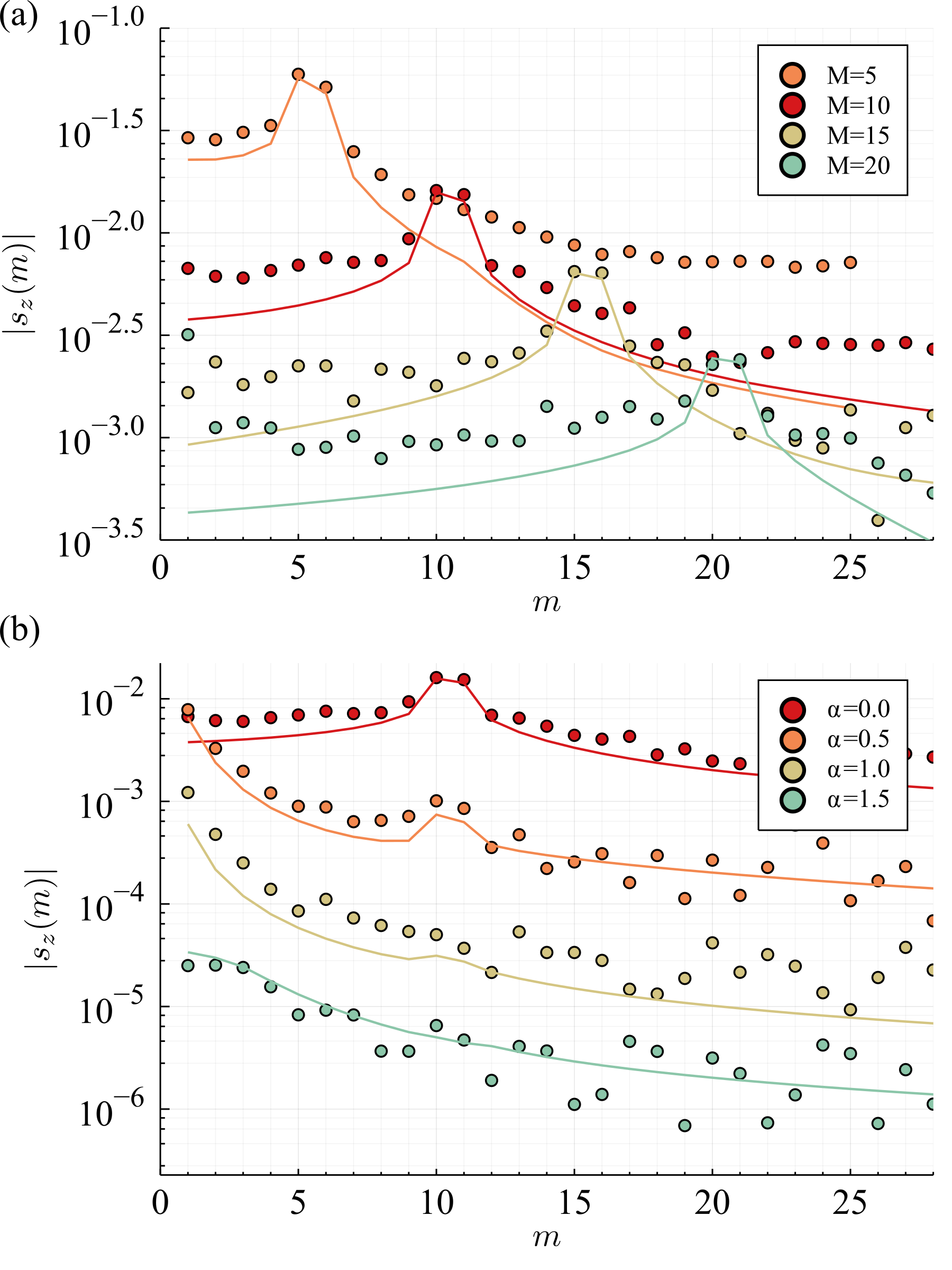}
	\caption[]{Amplitude of the discrete Fourier components $s_z(m)$: theory (lines) and experiment (dots). (a) For different values of $M$ at $\alpha=0$; (b) For different values of $\alpha$ at $M=10$. The pronounced peak at $m=M$ is an indicator of the long-range nature of the Floquet wavefunction (see text). For clarity, the curves with $\alpha>0$ in (b) are shifted vertically by $10^{2\alpha}$.}
	\label{fig:fourier_as}
\end{figure}

In the following, we propose and realize two complementary methods to experimentally probe the long-range nature of the model. The first method is based on the power-law scalings of the eigenstates of the enlarged Floquet Hamiltonian $\oh H_F$. In the absence of a cutoff, the Floquet functions are composed of a central peak at $m=0$ and long power-law tails,  $\abs{{\phi_n(m)}}^2\sim \abs{m}^{-x}$, see \cref{fig:alpha}(a). The value of $x$ can be analytically derived by the following scaling argument: If we consider the product of the $m$\textsuperscript{th} row of the matrix $\oh H_F$ with an eigenstate, the central peak gets multiplied by $h_x^{(m)}\sim \abs{m}^{-\alpha}$, while the long tail multiplies the $m$\textsuperscript{th} diagonal element of $\oh H_F$, giving rise to a contribution of the order $m^{1-x/2}$.
Because the eigenvalue does not depend on $m$, the two contributions needs to scale in the same way and, hence, $x=2+2\alpha$, or
\begin{align}
        \abs{{\phi_n(m)}}^2 \sim \frac1{\abs{m}^{2+2\alpha}}
	\label{eq:powerlaw}.
\end{align}
Note that for all $\alpha\ge 0$, one has $\sum_m \abs{\phi_n(m)}^2 < \infty$, ensuring that the Floquet functions $\ket{\phi_n(t)}$ are normalizable. In the presence of a cutoff $M$, the Floquet functions decay exponentially for $m>M$, see \cref{fig:alpha}(b). To preserve the normalization condition, the weight lost at $m \gg M$ appears as a pronounced peak at $m\approx M$.
{
This peak indicates that the Floquet functions are affected by the cutoff only at distances $m\sim M$ and appears sharp on a logarithmic scale.
Its integrated weight} is proportional to $\sum_{m>M} m^{-2-2\alpha} \approx 1/M^{1+2\alpha}$ and rapidly decreases with $\alpha$. This peak can be experimentally probed by studying the Fourier components $s_z(m)=T^{-1}\int_0^T \dd{t} s_z(t) e^{-i m\omega_p t}$, see \cref{fig:fourier_as}. The key feature that we observe is a pronounced peak at $m=M$. This peak is washed out as $\alpha$ increases and becomes invisible at $\alpha>1$, probing the scaling transition at $\alpha=1$.

\begin{figure}[t!]
	\centering	\includegraphics[width=.95\linewidth]{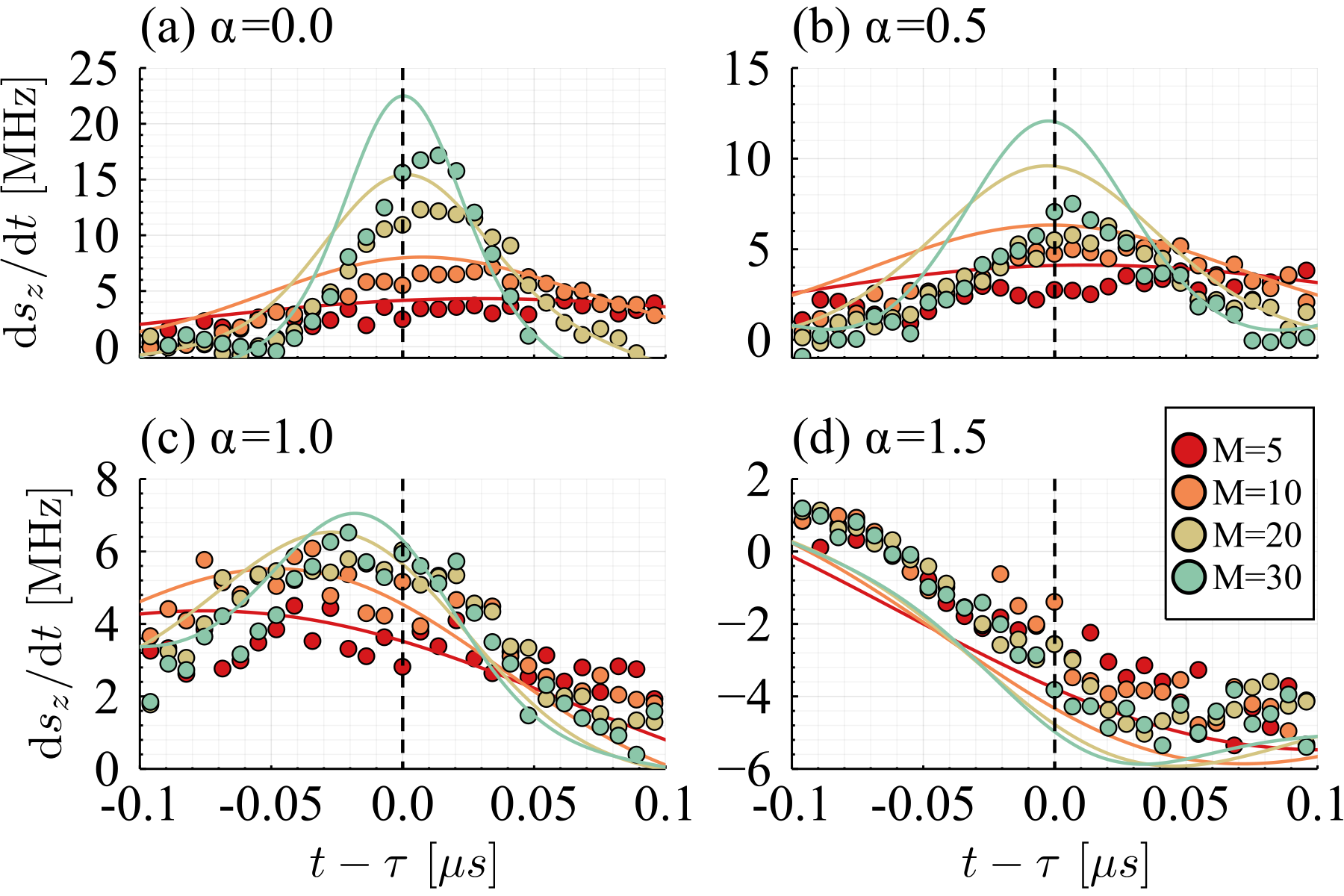}
        \caption{Time evolution of $\dv*{s_z}{t}$ in the vicinity of $t=\tau$: theory (lines) and experiment (dots). The cutoff parameter $M$ allows us to identify the transition between long-range and short-range hopping at $\alpha = 1$.}
	\label{fig:diff}
\end{figure}

Our second method to probe the scaling of the Floquet functions relies on the relation between the time derivatives of physical observables and the moments of $\ket{\phi_n(m)}$. According to \cref{eq:phi_dft}, the $q$\textsuperscript{th} time derivative of the expectation value of an operator $\oh O$ in the state $\ket{\phi_n(t)}$, $\pqty{\dv*{t}}^q\ev*{\oh O(t)}_n$, equals to
\begin{align*}
(i\omega_p)^q \sum_{m,m'}(m-m')^q\mel{\phi_n(m')}{\oh O}{\phi_n(m)}e^{i(m-m')\omega_p t}.
\end{align*}
A rigorous bound to this quantity can be obtained by assuming that $\oh O$ satisfies $\abs*{\mel*{\phi_n}{\oh O}{\phi_m}}\le \norm*{\oh O}$, where  $\norm*{\oh O}$ is the operator norm \footnote{Pauli matrices are examples of bounded operators with norm $\norm*{\oh O}=1$.}.
{This requirement is generically satisfied by operators that act on a finite Hilbert space, such as spin operators, and it does not apply to some operators of a infinite Hilbert space, such as the position or momentum operators of an oscillator, whose matrix elements can be arbitrary large.}
In this case,
\begin{align}
        \abs{\dv[q]{t}\ev*{\oh O(t)}} \le& \omega_p^q   \norm*{\oh O}
        \sum_{\mathrlap{m,m'}} \abs{m-m'}^{q} \braket{\phi_n(m)}{\phi_n(m')}.
	\label{eq:bound}
\end{align}
Using the scaling relation of \cref{eq:powerlaw}, {one can show that the right hand side of \cref{eq:bound} scales as $M^{q-\alpha}$ for large $M$, see appendix B.
This expression is finite for $q<\alpha$ and diverges for $q\ge\alpha$.}
In particular, for the kicked Hamiltonian ($\alpha=0$) the $q=0$ series diverges, highlighting the above-mentioned truncation problem. For all $\alpha \leq 1$, the $q=1$ series diverges, leading to a divergence of the time derivative $\dv*{t}\ev*{\oh O(t)}$. For larger values of $\alpha$, the $q=1$ series is convergent, but higher-order series (and, hence, higher order derivatives) are infinite. {Hence, in our model the time derivatives of physical observables can serve as order parameters of the scaling transition, in analogy to the role of two-point correlation functions for symmetry breaking phase transitions.} \Cref{fig:diff} compares the experimental measurement and numerical calculation of the time derivative of $s_z(t)$. The scaling behavior  changes drastically at the transition between long-range and short-range couplings ($\alpha=1$). For $\alpha\leq 1$, $\dv*{s_z}{t}\pqty{t=\tau}$ diverges with increasing $M$, while for $\alpha>1$ it saturates to a finite value, in agreement with our scaling argument {(see also appendix B)}.

In conclusion, we used Floquet engineering to simulate a one dimensional lattice with power-law hopping. We studied the scaling properties of the Floquet eigenstates and determined the effects of the long tails on the expectation values of physical observables and their time derivatives. By realizing this model on a quantum computer, we demonstrated the experimental capability of controlling and measuring a large number  ($M=30$) of harmonics. We were able to probe the long tails of the Floquet functions, both in real time and in Fourier space, and to study their dependence on the exponent $\alpha$. Our experimental observations probe the scaling transition between long-range and short-range couplings.

In recent years, superconducting circuits were used to simulate interesting \emph{nonequilibrium} effects \cite{bassman2021simulating,bharti2021noisy}, including many-body localization of photons \cite{roushan2017spectroscopic} and quench dynamics in one \cite{smith2019simulating} and two \cite{mei2020digital} dimensions. Our work adds an important tool to these simulators, namely pulse engineering for the realization of synthetic Floquet dimensions.
{
Pulse engineering allows us to both realize Floquet Hamiltonians with arbitrary couplings, and to measure the dynamics within each time period, the so called \emph{micromotion}.
By measuring the time harmonic decomposition of the Floquet function, and confirm their scaling properties.
Our approach goes beyond previous studies based on stroboscopic measurements, which probe the Floquet quasienergies, but are insensitive to the associated eigenfunctions.
}
By implementing our protocol on a multi-qubit quantum computer, it will be possible to explore equilibrium and nonequilibrium  phase transitions with long-range couplings.

\begin{acknowledgments} {\bfseries Data availability}	All MATLAB, Julia, Python, and QISKIT codes used to generate the experimental and theoretical data presented in this article are made freely available online in \cite{roses2021simulating}.

{\bfseries Acknowledgments} We thank Eli Arbel, Angelo Russomanno, and Alessandro Silva for useful discussions.
We thank the support  received through the IBM's QISKIT Slack channels.
MMR and EGDT are supported by the Israel Science Foundation grants No. 151/19 and 154/19.
We acknowledge use of the IBM Q for this work.
We acknowledge the access to advanced services provided by the IBM Quantum Researchers Program.
The views expressed are those of the authors, and do not reflect the official policy or position of IBM or the IBM Quantum team.
\end{acknowledgments}

\appendix
{
\section{Quasienergies crossings}
In this section we provide a brief analysis of the first Floquet-Brillouin zone of the periodically driven system with long-range interactions with the Floquet Hamiltonian (see \cref{eq:HF,eq:floquet_drive} in the main text), $\oh H_F$, and look for any spectral signatures related to the scaling transition. Because we are dealing with a two-level system, the folded spectrum has just two quasienergies. In the extended representation, this pair of quasienergies is repeated every integer multiple of the pump frequency, $\omega_p$.
Hence, we cannot observe a clustering of states, which is associated with a phase transition, but at most, a level crossing.

We numerically calculated the quasienergies in the first Floquet-Brillouin zone. Typical results of the quasienergies in the folded zone are shown in \cref{fig:app:der}(a)-(c), for different values of $h_0$. We, indeed, find that the quasienergies can cross leading to exceptional points of the system. These crossings generically do not occur at $\alpha=1$ and are not related to the transition between long-range and short range couplings discussed in this work.

\begin{figure}[b]
\centering
\includegraphics[width=\linewidth]{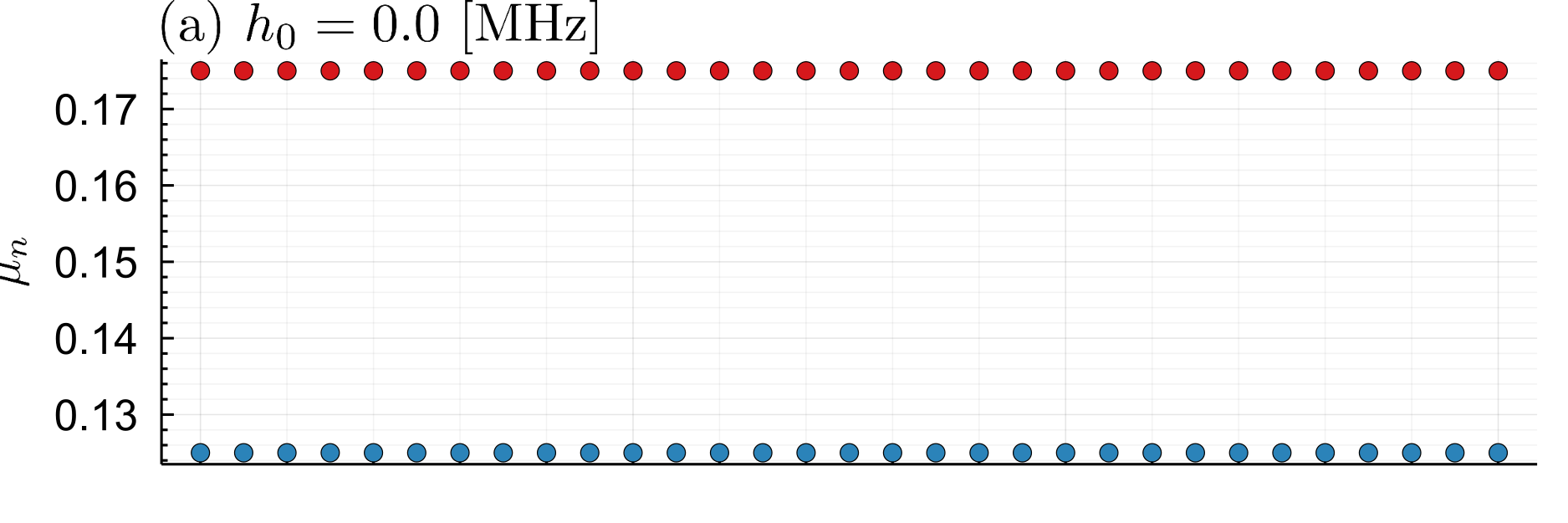}\\
\includegraphics[width=\linewidth]{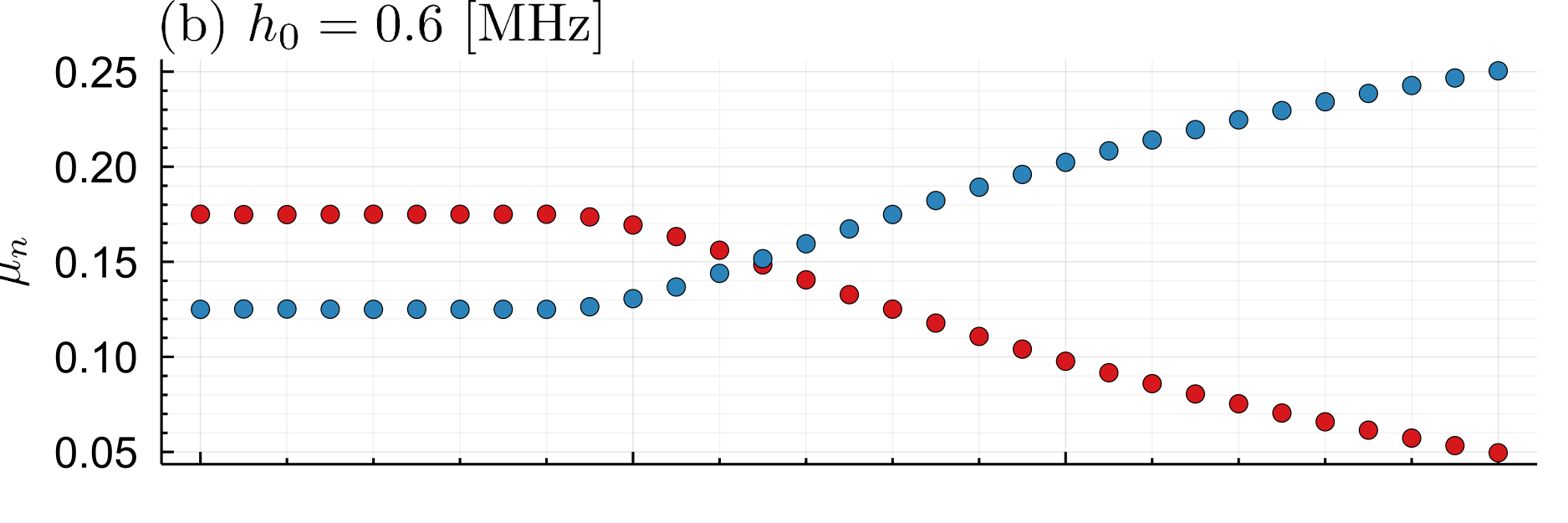}\\
\includegraphics[width=\linewidth]{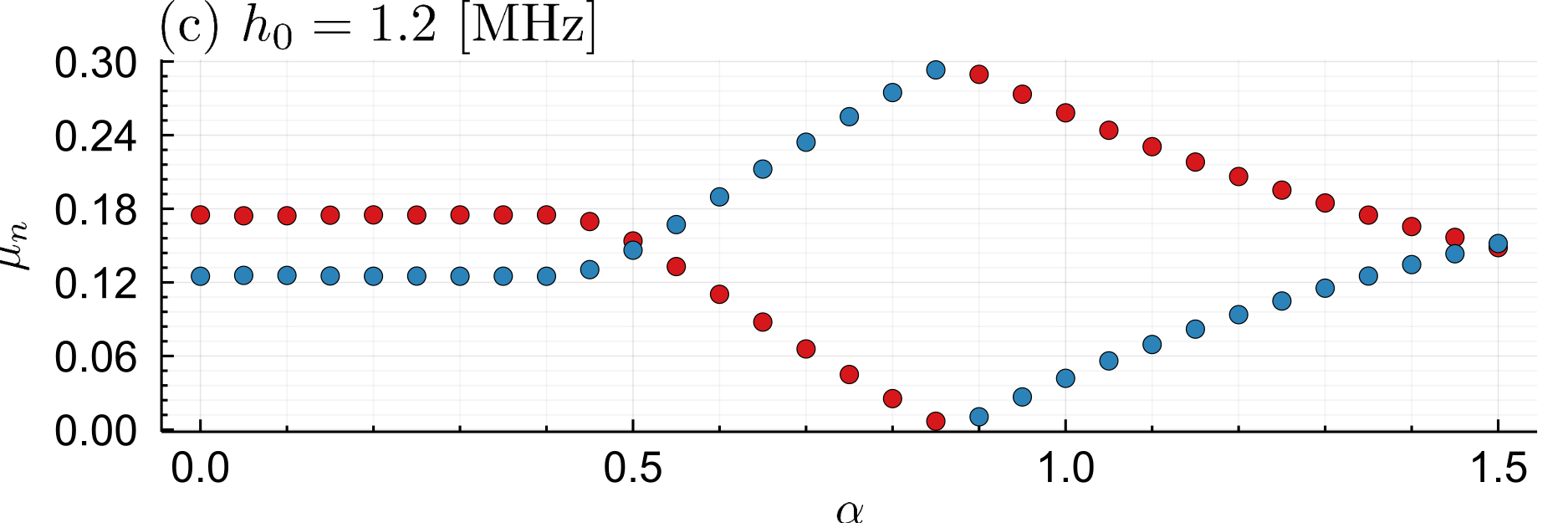}
\caption{The quasienergies of \cref{eq:HF,eq:powerlaw} for various values of the driving strength in the first Floquet-Brillouin zone. Each color represents a different quasienergy.}
\label{fig:a:quasi}
\end{figure}

\section{Scaling study of the first derivative of $s_z(t)$}
This section provides a deeper analysis of theoretical predictions related to \cref{fig:diff} in the main text. In \cref{eq:bound}, we have obtained an upper bound for any derivative of generic observables\footnote{For the sake of brevity we limit our operators to those that have a norm of 1, \emph{i.e.\@}, $\norm*{\oh O}=1$}, $\oh O(t)$.
We will now show how to use the scaling law obtained in \cref{eq:powerlaw} in \cref{eq:bound}.
We begin by remembering that the eigenfunctions scales as
\begin{align}
\label{eq:app:scale-not}
\abs{\ket{\phi_n(m\neq0)}}\sim&\abs{m}^{-1-\alpha},\\
\label{eq:app:scale-yes}
\abs{\ket{\phi_n(m=0)}}\sim&\mathcal{O}(1)
\end{align}
We now separate the sum over $m$ and $m'$ in \cref{eq:bound} to two sums, one in which $m$ or $m'$ equals zero and another where $m\neq0$ and $m'\neq0$
\begin{align}
\label{eq:app:sums}
\sum\limits_{m,m'}\to&\sum\limits_{\substack{m,m'\\m\neq0,m'\neq0}}+\sum\limits_{\substack{m\\m'=0}},
\end{align}
We, first, consider the double sum with no zeros.
For a large cutoff, $M$, we can approximate the double sum to a double integral
\begin{align}
\sum\limits_{\substack{m,m'\\m\neq0,m'\neq0}}\to&\iint\limits_{\substack{m\neq0\\m'\neq0}}\dd{m}\dd{m'},
\end{align}
using the scaling law of \cref{eq:app:scale-not} we can rewrite \cref{eq:bound} as
\begin{multline}
\abs{\dv[q]{t}\ev*{\oh O(t)}}\le\omega_p^q\norm*{\oh O}
\\\times
\iint\limits_{\substack{m\neq0\\m'\neq0}}\dd{m}\dd{m'}\abs{m-m'}^q\abs{m}^{-1-\alpha}\abs{m'}^{-1-\alpha},
\end{multline}
we will examine the scaling of the integral, so we will ignore any numerical constants for the sake of brevity
\begin{multline}
\iint\limits_{\substack{m\neq0\\m'\neq0}}\dd{m}\dd{m'}\abs{m-m'}^q\abs{m}^{-1-\alpha}\abs{m'}^{-1-\alpha}
\\\begin{aligned}
\\\sim&
\int\limits_1^M\dd{m}\int\limits_1^m\dd{m'}\pqty{m-m'}^qm^{-1-\alpha}\pqty{m'}^{-1-\alpha}
\\\sim&
\int\limits_1^M\dd{m}\int\limits_1^m\dd{m'}\sum\limits_{k=0}^qm^{q-k}\pqty{m'}^km^{-1-\alpha}\pqty{m'}^{-1-\alpha}
\\\sim&
\int\limits_1^M\dd{m}\int\limits_1^m\dd{m'}\sum\limits_{k=0}^q
m^{q-k-1-\alpha}\pqty{m'}^{k-1-\alpha}
\\\sim&
\int\limits_1^M\dd{m}\sum\limits_{k=0}^qm^{q-k-1-\alpha}m^{k-\alpha}
\\\sim&
\int\limits_1^M\dd{m}m^{q-1-2\alpha}.
\end{aligned}
\end{multline}
At this point we need to preform the integral carefully.
If $q=2\alpha$ the integral has a logarithmic behavior.
Otherwise, if $q>2\alpha$, the integral scales as $M^{q-2\alpha}$ or saturates to a finite values of order ine if $q<2\alpha$.
Hence, we find
\begin{multline}
\label{eq:app:non}
\iint\limits_{\substack{m\neq0\\m'\neq0}}\dd{m}\dd{m'}\abs{m-m'}^q\abs{m}^{-1-\alpha}\abs{m'}^{-1-\alpha}\\\sim
\begin{cases}
\ln(M),&q=2\alpha,\\
M^{q-2\alpha},&q>2\alpha,\\
1,&q\le2\alpha.
\end{cases}.
\end{multline}

Now we will focus on the sum in which either $m$ or $m'$ equals zero.
As before, we convert the sum into an integral
\begin{align}
\sum\limits_{\substack{m\\m'=0}}\to&\int\limits_1^M\dd{m},\\
\end{align}
plugging in the scaling law of \cref{eq:app:scale-yes} we rewrite \cref{eq:bound}
\begin{align}
\begin{split}
\int\limits_1^M\dd{m}\abs{m}^{-1-\alpha}\abs{m}^q\sim&\int\limits_1^M\dd{m}\pqty{m}^{q-1-\alpha}.
\end{split}
\end{align}
Just like before, we need to preform the integral carefully.
If $q=\alpha$ the integral has a logarithmic behavior.
Otherwise, if $q>\alpha$ the integral scales as $M^{q-\alpha}$, or scales as 1 if $q<\alpha$.
\begin{align}
\label{eq:app:with}
\int\limits_1^M\dd{m}\abs{m}^{-1-\alpha}\abs{m}^q\sim&
\begin{cases}
\ln(M),&q=\alpha,\\
M^{q-\alpha},&q>\alpha,\\
1,&q\le\alpha.
\end{cases}.
\end{align}

Using \cref{eq:app:sums,eq:app:non,eq:app:with} we can deduce the leading behavior of \cref{eq:bound} presented in the text
\begin{align}
        \label{eq:app:bound}
        \abs{\dv[]{t}\ev*{\oh O(t)}}\le &
        \begin{cases}
        M^{1-\alpha},&\alpha<1,\\
        \ln(M),&\alpha=1,\\
        \mathcal{O}(1),&\alpha>1.
        \end{cases}.
\end{align}

Note that this bound can be alternatively derived using the Heisenberg equation of motion, $\dv*{t} \ev*{\oh O(t)}=i\ev*{\comm*{\oh H}{\oh O}}/\hbar$, showing that the time derivative of physical operators is linearly proportional to the Hamiltonian and, hence, to $\abs{h_x(t)}\le \sum_m \abs*{h_x^{(m)}}$. This quantity converges for all $\alpha>1$, and it can diverge only for $\alpha\le 1$.

Due to physical limitations, in the experiment we were only able to realize our model only up to $M=30$ (with accurate results). To accurately describe the scaling transition, we plot the theoretical value of $\abs{\dv*{s_z}{t}}$, as a function of $M$ in Fig.~\ref{fig:app:der}. As expected, we observe that this quantity diverges as $M^{1-\alpha}$ for $\alpha<1$ and remains finite for $\alpha>1$. At the scaling transition, for $\alpha=1$, the time derivative diverges logarithmically.

\begin{figure}[ht]
        \includegraphics[width=\linewidth]{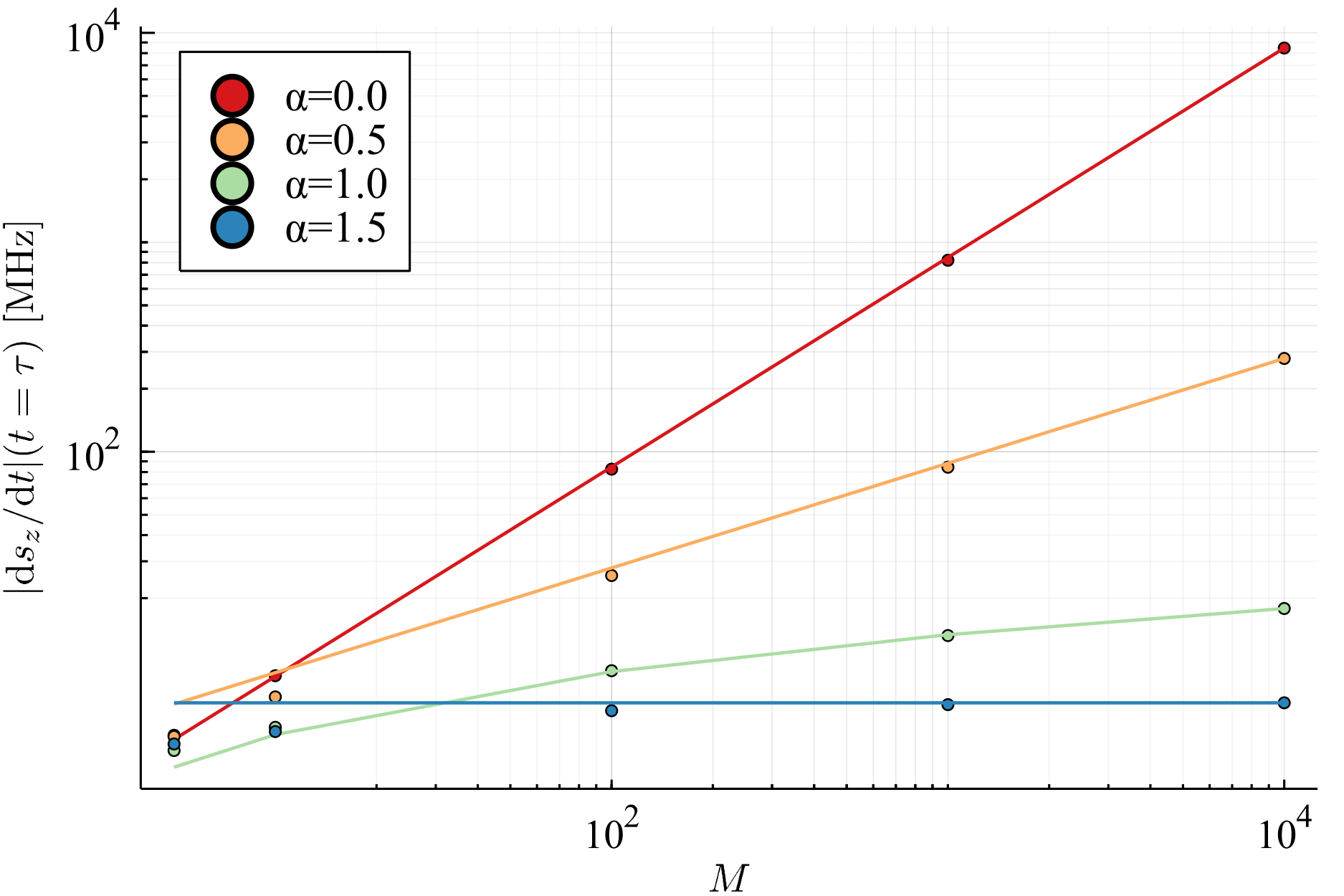}
        \caption{Time derivative $\abs{\dv*{s_z}{t}}$ at $t=\tau$ as a function of the cut-off parameter, $M$. This plot allows us to identify a transition between long-range and short-range hopping at $\alpha=1$. The dots are simulated data and the solid lines show the expected scaling behavior, \cref{eq:app:bound}.}
        \label{fig:app:der}
\end{figure}

}

\begin{thebibliography}{63}%
\makeatletter
\providecommand \@ifxundefined [1]{%
 \@ifx{#1\undefined}
}%
\providecommand \@ifnum [1]{%
 \ifnum #1\expandafter \@firstoftwo
 \else \expandafter \@secondoftwo
 \fi
}%
\providecommand \@ifx [1]{%
 \ifx #1\expandafter \@firstoftwo
 \else \expandafter \@secondoftwo
 \fi
}%
\providecommand \natexlab [1]{#1}%
\providecommand \enquote  [1]{``#1''}%
\providecommand \bibnamefont  [1]{#1}%
\providecommand \bibfnamefont [1]{#1}%
\providecommand \citenamefont [1]{#1}%
\providecommand \href@noop [0]{\@secondoftwo}%
\providecommand \href [0]{\begingroup \@sanitize@url \@href}%
\providecommand \@href[1]{\@@startlink{#1}\@@href}%
\providecommand \@@href[1]{\endgroup#1\@@endlink}%
\providecommand \@sanitize@url [0]{\catcode `\\12\catcode `\$12\catcode
  `\&12\catcode `\#12\catcode `\^12\catcode `\_12\catcode `\%12\relax}%
\providecommand \@@startlink[1]{}%
\providecommand \@@endlink[0]{}%
\providecommand \url  [0]{\begingroup\@sanitize@url \@url }%
\providecommand \@url [1]{\endgroup\@href {#1}{\urlprefix }}%
\providecommand \urlprefix  [0]{URL }%
\providecommand \Eprint [0]{\href }%
\providecommand \doibase [0]{https://doi.org/}%
\providecommand \selectlanguage [0]{\@gobble}%
\providecommand \bibinfo  [0]{\@secondoftwo}%
\providecommand \bibfield  [0]{\@secondoftwo}%
\providecommand \translation [1]{[#1]}%
\providecommand \BibitemOpen [0]{}%
\providecommand \bibitemStop [0]{}%
\providecommand \bibitemNoStop [0]{.\EOS\space}%
\providecommand \EOS [0]{\spacefactor3000\relax}%
\providecommand \BibitemShut  [1]{\csname bibitem#1\endcsname}%
\let\auto@bib@innerbib\@empty
\bibitem [{\citenamefont {Morigi}\ and\ \citenamefont
  {Fishman}(2004)}]{morigi2004eigenmodes}%
  \BibitemOpen
  \bibfield  {author} {\bibinfo {author} {\bibfnamefont {G.}~\bibnamefont
  {Morigi}}\ and\ \bibinfo {author} {\bibfnamefont {S.}~\bibnamefont
  {Fishman}},\ }\bibfield  {title} {\bibinfo {title} {Eigenmodes and
  thermodynamics of a coulomb chain in a harmonic potential},\ }\href
  {https://doi.org/10.1103/PhysRevLett.93.170602} {\bibfield  {journal}
  {\bibinfo  {journal} {Physical Review Letters}\ }\textbf {\bibinfo {volume}
  {93}},\ \bibinfo {pages} {170602} (\bibinfo {year} {2004})}\BibitemShut
  {NoStop}%
\bibitem [{\citenamefont {Bouchet}\ \emph {et~al.}(2010)\citenamefont
  {Bouchet}, \citenamefont {Gupta},\ and\ \citenamefont
  {Mukamel}}]{bouchet2010thermodynamics}%
  \BibitemOpen
  \bibfield  {author} {\bibinfo {author} {\bibfnamefont {F.}~\bibnamefont
  {Bouchet}}, \bibinfo {author} {\bibfnamefont {S.}~\bibnamefont {Gupta}},\
  and\ \bibinfo {author} {\bibfnamefont {D.}~\bibnamefont {Mukamel}},\
  }\bibfield  {title} {\bibinfo {title} {Thermodynamics and dynamics of systems
  with long-range interactions},\ }\href
  {https://doi.org/https://doi.org/10.1016/j.physa.2010.02.024} {\bibfield
  {journal} {\bibinfo  {journal} {Physica A: Statistical Mechanics and its
  Applications}\ }\textbf {\bibinfo {volume} {389}},\ \bibinfo {pages} {4389 }
  (\bibinfo {year} {2010})},\ \bibinfo {note} {proceedings of the 12th
  International Summer School on Fundamental Problems in Statistical
  Physics}\BibitemShut {NoStop}%
\bibitem [{\citenamefont {Gupta}\ and\ \citenamefont
  {Ruffo}(2017)}]{gupta2017world}%
  \BibitemOpen
  \bibfield  {author} {\bibinfo {author} {\bibfnamefont {S.}~\bibnamefont
  {Gupta}}\ and\ \bibinfo {author} {\bibfnamefont {S.}~\bibnamefont {Ruffo}},\
  }\bibfield  {title} {\bibinfo {title} {The world of long-range interactions:
  A bird’s eye view},\ }\href {https://doi.org/10.1142/S0217751X17410184}
  {\bibfield  {journal} {\bibinfo  {journal} {International Journal of Modern
  Physics A}\ }\textbf {\bibinfo {volume} {32}},\ \bibinfo {pages} {1741018}
  (\bibinfo {year} {2017})}\BibitemShut {NoStop}%
\bibitem [{\citenamefont {Campa}\ \emph {et~al.}(2009)\citenamefont {Campa},
  \citenamefont {Dauxois},\ and\ \citenamefont {Ruffo}}]{campa2009statistical}%
  \BibitemOpen
  \bibfield  {author} {\bibinfo {author} {\bibfnamefont {A.}~\bibnamefont
  {Campa}}, \bibinfo {author} {\bibfnamefont {T.}~\bibnamefont {Dauxois}},\
  and\ \bibinfo {author} {\bibfnamefont {S.}~\bibnamefont {Ruffo}},\ }\bibfield
   {title} {\bibinfo {title} {Statistical mechanics and dynamics of solvable
  models with long-range interactions},\ }\href
  {https://doi.org/https://doi.org/10.1016/j.physrep.2009.07.001} {\bibfield
  {journal} {\bibinfo  {journal} {Physics Reports}\ }\textbf {\bibinfo {volume}
  {480}},\ \bibinfo {pages} {57} (\bibinfo {year} {2009})}\BibitemShut
  {NoStop}%
\bibitem [{\citenamefont {Campa}\ \emph {et~al.}(2014)\citenamefont {Campa},
  \citenamefont {Dauxois}, \citenamefont {Fanelli},\ and\ \citenamefont
  {Ruffo}}]{campa2014physics}%
  \BibitemOpen
  \bibfield  {author} {\bibinfo {author} {\bibfnamefont {A.}~\bibnamefont
  {Campa}}, \bibinfo {author} {\bibfnamefont {T.}~\bibnamefont {Dauxois}},
  \bibinfo {author} {\bibfnamefont {D.}~\bibnamefont {Fanelli}},\ and\ \bibinfo
  {author} {\bibfnamefont {S.}~\bibnamefont {Ruffo}},\ }\href
  {https://doi.org/10.1093/acprof:oso/9780199581931.001.0001} {\emph {\bibinfo
  {title} {Physics of long-range interacting systems}}}\ (\bibinfo  {publisher}
  {OUP Oxford},\ \bibinfo {year} {2014})\BibitemShut {NoStop}%
\bibitem [{\citenamefont {Richerme}\ \emph {et~al.}(2014)\citenamefont
  {Richerme}, \citenamefont {Gong}, \citenamefont {Lee}, \citenamefont {Senko},
  \citenamefont {Smith}, \citenamefont {Foss-Feig}, \citenamefont {Michalakis},
  \citenamefont {Gorshkov},\ and\ \citenamefont {Monroe}}]{richerme2014non}%
  \BibitemOpen
  \bibfield  {author} {\bibinfo {author} {\bibfnamefont {P.}~\bibnamefont
  {Richerme}}, \bibinfo {author} {\bibfnamefont {Z.-X.}\ \bibnamefont {Gong}},
  \bibinfo {author} {\bibfnamefont {A.}~\bibnamefont {Lee}}, \bibinfo {author}
  {\bibfnamefont {C.}~\bibnamefont {Senko}}, \bibinfo {author} {\bibfnamefont
  {J.}~\bibnamefont {Smith}}, \bibinfo {author} {\bibfnamefont
  {M.}~\bibnamefont {Foss-Feig}}, \bibinfo {author} {\bibfnamefont
  {S.}~\bibnamefont {Michalakis}}, \bibinfo {author} {\bibfnamefont {A.~V.}\
  \bibnamefont {Gorshkov}},\ and\ \bibinfo {author} {\bibfnamefont
  {C.}~\bibnamefont {Monroe}},\ }\bibfield  {title} {\bibinfo {title}
  {Non-local propagation of correlations in quantum systems with long-range
  interactions},\ }\href {https://doi.org/10.1038/nature13450} {\bibfield
  {journal} {\bibinfo  {journal} {Nature}\ }\textbf {\bibinfo {volume} {511}},\
  \bibinfo {pages} {198} (\bibinfo {year} {2014})}\BibitemShut {NoStop}%
\bibitem [{\citenamefont {Zhang}\ \emph {et~al.}(2017)\citenamefont {Zhang},
  \citenamefont {Pagano}, \citenamefont {Hess}, \citenamefont {Kyprianidis},
  \citenamefont {Becker}, \citenamefont {Kaplan}, \citenamefont {Gorshkov},
  \citenamefont {Gong},\ and\ \citenamefont {Monroe}}]{zhang2017observation}%
  \BibitemOpen
  \bibfield  {author} {\bibinfo {author} {\bibfnamefont {J.}~\bibnamefont
  {Zhang}}, \bibinfo {author} {\bibfnamefont {G.}~\bibnamefont {Pagano}},
  \bibinfo {author} {\bibfnamefont {P.~W.}\ \bibnamefont {Hess}}, \bibinfo
  {author} {\bibfnamefont {A.}~\bibnamefont {Kyprianidis}}, \bibinfo {author}
  {\bibfnamefont {P.}~\bibnamefont {Becker}}, \bibinfo {author} {\bibfnamefont
  {H.}~\bibnamefont {Kaplan}}, \bibinfo {author} {\bibfnamefont {A.~V.}\
  \bibnamefont {Gorshkov}}, \bibinfo {author} {\bibfnamefont {Z.-X.}\
  \bibnamefont {Gong}},\ and\ \bibinfo {author} {\bibfnamefont
  {C.}~\bibnamefont {Monroe}},\ }\bibfield  {title} {\bibinfo {title}
  {Observation of a many-body dynamical phase transition with a 53-qubit
  quantum simulator},\ }\href {https://doi.org/10.1038/nature24654} {\bibfield
  {journal} {\bibinfo  {journal} {Nature}\ }\textbf {\bibinfo {volume} {551}},\
  \bibinfo {pages} {601} (\bibinfo {year} {2017})}\BibitemShut {NoStop}%
\bibitem [{\citenamefont {\ifmmode \check{Z}\else
  \v{Z}\fi{}unkovi\ifmmode~\check{c}\else \v{c}\fi{}}\ \emph
  {et~al.}(2018)\citenamefont {\ifmmode \check{Z}\else
  \v{Z}\fi{}unkovi\ifmmode~\check{c}\else \v{c}\fi{}}, \citenamefont {Heyl},
  \citenamefont {Knap},\ and\ \citenamefont {Silva}}]{vzunkovivc2018dynamical}%
  \BibitemOpen
  \bibfield  {author} {\bibinfo {author} {\bibfnamefont {B.}~\bibnamefont
  {\ifmmode \check{Z}\else \v{Z}\fi{}unkovi\ifmmode~\check{c}\else
  \v{c}\fi{}}}, \bibinfo {author} {\bibfnamefont {M.}~\bibnamefont {Heyl}},
  \bibinfo {author} {\bibfnamefont {M.}~\bibnamefont {Knap}},\ and\ \bibinfo
  {author} {\bibfnamefont {A.}~\bibnamefont {Silva}},\ }\bibfield  {title}
  {\bibinfo {title} {Dynamical quantum phase transitions in spin chains with
  long-range interactions: Merging different concepts of nonequilibrium
  criticality},\ }\href {https://doi.org/10.1103/PhysRevLett.120.130601}
  {\bibfield  {journal} {\bibinfo  {journal} {Physical Review Letters}\
  }\textbf {\bibinfo {volume} {120}},\ \bibinfo {pages} {130601} (\bibinfo
  {year} {2018})}\BibitemShut {NoStop}%
\bibitem [{\citenamefont {Dyson}(1969)}]{dyson1969existence}%
  \BibitemOpen
  \bibfield  {author} {\bibinfo {author} {\bibfnamefont {F.~J.}\ \bibnamefont
  {Dyson}},\ }\bibfield  {title} {\bibinfo {title} {Existence of a
  phase-transition in a one-dimensional ising ferromagnet},\ }\href
  {https://doi.org/10.1007/BF01645907} {\bibfield  {journal} {\bibinfo
  {journal} {Communications in Mathematical Physics}\ }\textbf {\bibinfo
  {volume} {12}},\ \bibinfo {pages} {91} (\bibinfo {year} {1969})}\BibitemShut
  {NoStop}%
\bibitem [{\citenamefont {Thouless}(1969)}]{thouless1969long}%
  \BibitemOpen
  \bibfield  {author} {\bibinfo {author} {\bibfnamefont {D.}~\bibnamefont
  {Thouless}},\ }\bibfield  {title} {\bibinfo {title} {Long-range order in
  one-dimensional ising systems},\ }\href
  {https://doi.org/10.1103/PhysRev.187.732} {\bibfield  {journal} {\bibinfo
  {journal} {Physical Review}\ }\textbf {\bibinfo {volume} {187}},\ \bibinfo
  {pages} {732} (\bibinfo {year} {1969})}\BibitemShut {NoStop}%
\bibitem [{\citenamefont {Sak}(1973)}]{sak1973recursion}%
  \BibitemOpen
  \bibfield  {author} {\bibinfo {author} {\bibfnamefont {J.}~\bibnamefont
  {Sak}},\ }\bibfield  {title} {\bibinfo {title} {Recursion relations and fixed
  points for ferromagnets with long-range interactions},\ }\href
  {https://doi.org/10.1103/PhysRevB.8.281} {\bibfield  {journal} {\bibinfo
  {journal} {Physical Review B}\ }\textbf {\bibinfo {volume} {8}},\ \bibinfo
  {pages} {281} (\bibinfo {year} {1973})}\BibitemShut {NoStop}%
\bibitem [{\citenamefont {Levitov}(1999)}]{levitov1999critical}%
  \BibitemOpen
  \bibfield  {author} {\bibinfo {author} {\bibfnamefont {L.}~\bibnamefont
  {Levitov}},\ }\bibfield  {title} {\bibinfo {title} {Critical hamiltonians
  with long range hopping},\ }\href
  {https://doi.org/10.1002/(SICI)1521-3889(199911)8:7/9<697::AID-ANDP697>3.0.CO;2-W}
  {\bibfield  {journal} {\bibinfo  {journal} {Annalen der Physik}\ }\textbf
  {\bibinfo {volume} {8}},\ \bibinfo {pages} {697} (\bibinfo {year}
  {1999})}\BibitemShut {NoStop}%
\bibitem [{\citenamefont {Nag}\ and\ \citenamefont {Garg}(2019)}]{nag2019many}%
  \BibitemOpen
  \bibfield  {author} {\bibinfo {author} {\bibfnamefont {S.}~\bibnamefont
  {Nag}}\ and\ \bibinfo {author} {\bibfnamefont {A.}~\bibnamefont {Garg}},\
  }\bibfield  {title} {\bibinfo {title} {Many-body localization in the presence
  of long-range interactions and long-range hopping},\ }\href
  {https://doi.org/10.1103/PhysRevB.99.224203} {\bibfield  {journal} {\bibinfo
  {journal} {Physical Review B}\ }\textbf {\bibinfo {volume} {99}},\ \bibinfo
  {pages} {224203} (\bibinfo {year} {2019})}\BibitemShut {NoStop}%
\bibitem [{\citenamefont {Brezin}\ \emph {et~al.}(2014)\citenamefont {Brezin},
  \citenamefont {Parisi},\ and\ \citenamefont
  {Ricci-Tersenghi}}]{brezin2014crossover}%
  \BibitemOpen
  \bibfield  {author} {\bibinfo {author} {\bibfnamefont {E.}~\bibnamefont
  {Brezin}}, \bibinfo {author} {\bibfnamefont {G.}~\bibnamefont {Parisi}},\
  and\ \bibinfo {author} {\bibfnamefont {F.}~\bibnamefont {Ricci-Tersenghi}},\
  }\bibfield  {title} {\bibinfo {title} {The crossover region between
  long-range and short-range interactions for the critical exponents},\ }\href
  {https://doi.org/10.1007/s10955-014-1081-0} {\bibfield  {journal} {\bibinfo
  {journal} {Journal of Statistical Physics}\ }\textbf {\bibinfo {volume}
  {157}},\ \bibinfo {pages} {855} (\bibinfo {year} {2014})}\BibitemShut
  {NoStop}%
\bibitem [{\citenamefont {Behan}\ \emph {et~al.}(2017)\citenamefont {Behan},
  \citenamefont {Rastelli}, \citenamefont {Rychkov},\ and\ \citenamefont
  {Zan}}]{behan2017long}%
  \BibitemOpen
  \bibfield  {author} {\bibinfo {author} {\bibfnamefont {C.}~\bibnamefont
  {Behan}}, \bibinfo {author} {\bibfnamefont {L.}~\bibnamefont {Rastelli}},
  \bibinfo {author} {\bibfnamefont {S.}~\bibnamefont {Rychkov}},\ and\ \bibinfo
  {author} {\bibfnamefont {B.}~\bibnamefont {Zan}},\ }\bibfield  {title}
  {\bibinfo {title} {Long-range critical exponents near the short-range
  crossover},\ }\href {https://doi.org/10.1103/PhysRevLett.118.241601}
  {\bibfield  {journal} {\bibinfo  {journal} {Physical Review Letters}\
  }\textbf {\bibinfo {volume} {118}},\ \bibinfo {pages} {241601} (\bibinfo
  {year} {2017})}\BibitemShut {NoStop}%
\bibitem [{\citenamefont {Defenu}\ \emph {et~al.}(2020)\citenamefont {Defenu},
  \citenamefont {Codello}, \citenamefont {Ruffo},\ and\ \citenamefont
  {Trombettoni}}]{defenu2020criticality}%
  \BibitemOpen
  \bibfield  {author} {\bibinfo {author} {\bibfnamefont {N.}~\bibnamefont
  {Defenu}}, \bibinfo {author} {\bibfnamefont {A.}~\bibnamefont {Codello}},
  \bibinfo {author} {\bibfnamefont {S.}~\bibnamefont {Ruffo}},\ and\ \bibinfo
  {author} {\bibfnamefont {A.}~\bibnamefont {Trombettoni}},\ }\bibfield
  {title} {\bibinfo {title} {Criticality of spin systems with weak long-range
  interactions},\ }\href {https://doi.org/10.1088/1751-8121/ab6a6c} {\bibfield
  {journal} {\bibinfo  {journal} {Journal of Physics A: Mathematical and
  Theoretical}\ }\textbf {\bibinfo {volume} {53}},\ \bibinfo {pages} {143001}
  (\bibinfo {year} {2020})}\BibitemShut {NoStop}%
\bibitem [{\citenamefont {Campa}\ \emph {et~al.}(2019)\citenamefont {Campa},
  \citenamefont {Gori}, \citenamefont {Hovhannisyan}, \citenamefont {Ruffo},\
  and\ \citenamefont {Trombettoni}}]{campa2019ising}%
  \BibitemOpen
  \bibfield  {author} {\bibinfo {author} {\bibfnamefont {A.}~\bibnamefont
  {Campa}}, \bibinfo {author} {\bibfnamefont {G.}~\bibnamefont {Gori}},
  \bibinfo {author} {\bibfnamefont {V.}~\bibnamefont {Hovhannisyan}}, \bibinfo
  {author} {\bibfnamefont {S.}~\bibnamefont {Ruffo}},\ and\ \bibinfo {author}
  {\bibfnamefont {A.}~\bibnamefont {Trombettoni}},\ }\bibfield  {title}
  {\bibinfo {title} {Ising chains with competing interactions in the presence
  of long-range couplings},\ }\href {https://doi.org/10.1088/1751-8121/ab2baf}
  {\bibfield  {journal} {\bibinfo  {journal} {Journal of Physics A:
  Mathematical and Theoretical}\ }\textbf {\bibinfo {volume} {52}},\ \bibinfo
  {pages} {344002} (\bibinfo {year} {2019})}\BibitemShut {NoStop}%
\bibitem [{\citenamefont {Xiao}\ \emph {et~al.}(2019)\citenamefont {Xiao},
  \citenamefont {H{\'e}bert}, \citenamefont {Batrouni},\ and\ \citenamefont
  {Scalettar}}]{xiao2019competition}%
  \BibitemOpen
  \bibfield  {author} {\bibinfo {author} {\bibfnamefont {B.}~\bibnamefont
  {Xiao}}, \bibinfo {author} {\bibfnamefont {F.}~\bibnamefont {H{\'e}bert}},
  \bibinfo {author} {\bibfnamefont {G.}~\bibnamefont {Batrouni}},\ and\
  \bibinfo {author} {\bibfnamefont {R.}~\bibnamefont {Scalettar}},\ }\bibfield
  {title} {\bibinfo {title} {Competition between phase separation and spin
  density wave or charge density wave order: Role of long-range interactions},\
  }\href {https://doi.org/10.1103/PhysRevB.99.205145} {\bibfield  {journal}
  {\bibinfo  {journal} {Physical Review B}\ }\textbf {\bibinfo {volume} {99}},\
  \bibinfo {pages} {205145} (\bibinfo {year} {2019})}\BibitemShut {NoStop}%
\bibitem [{\citenamefont {Porras}\ and\ \citenamefont
  {Cirac}(2004)}]{porras2004effective}%
  \BibitemOpen
  \bibfield  {author} {\bibinfo {author} {\bibfnamefont {D.}~\bibnamefont
  {Porras}}\ and\ \bibinfo {author} {\bibfnamefont {J.~I.}\ \bibnamefont
  {Cirac}},\ }\bibfield  {title} {\bibinfo {title} {Effective quantum spin
  systems with trapped ions},\ }\href
  {https://doi.org/10.1103/PhysRevLett.92.207901} {\bibfield  {journal}
  {\bibinfo  {journal} {Physical Review Letters}\ }\textbf {\bibinfo {volume}
  {92}},\ \bibinfo {pages} {207901} (\bibinfo {year} {2004})}\BibitemShut
  {NoStop}%
\bibitem [{\citenamefont {Islam}\ \emph {et~al.}(2013)\citenamefont {Islam},
  \citenamefont {Senko}, \citenamefont {Campbell}, \citenamefont {Korenblit},
  \citenamefont {Smith}, \citenamefont {Lee}, \citenamefont {Edwards},
  \citenamefont {Wang}, \citenamefont {Freericks},\ and\ \citenamefont
  {Monroe}}]{islam2013emergence}%
  \BibitemOpen
  \bibfield  {author} {\bibinfo {author} {\bibfnamefont {R.}~\bibnamefont
  {Islam}}, \bibinfo {author} {\bibfnamefont {C.}~\bibnamefont {Senko}},
  \bibinfo {author} {\bibfnamefont {W.}~\bibnamefont {Campbell}}, \bibinfo
  {author} {\bibfnamefont {S.}~\bibnamefont {Korenblit}}, \bibinfo {author}
  {\bibfnamefont {J.}~\bibnamefont {Smith}}, \bibinfo {author} {\bibfnamefont
  {A.}~\bibnamefont {Lee}}, \bibinfo {author} {\bibfnamefont {E.}~\bibnamefont
  {Edwards}}, \bibinfo {author} {\bibfnamefont {C.-C.}\ \bibnamefont {Wang}},
  \bibinfo {author} {\bibfnamefont {J.}~\bibnamefont {Freericks}},\ and\
  \bibinfo {author} {\bibfnamefont {C.}~\bibnamefont {Monroe}},\ }\bibfield
  {title} {\bibinfo {title} {Emergence and frustration of magnetism with
  variable-range interactions in a quantum simulator},\ }\href
  {https://doi.org/10.1126/science.1232296} {\bibfield  {journal} {\bibinfo
  {journal} {Science}\ }\textbf {\bibinfo {volume} {340}},\ \bibinfo {pages}
  {583} (\bibinfo {year} {2013})}\BibitemShut {NoStop}%
\bibitem [{Note1()}]{Note1}%
  \BibitemOpen
  \bibinfo {note} {See Ref.~\cite {rudner2020floquet} for an
  introduction}\BibitemShut {NoStop}%
\bibitem [{\citenamefont {Cross}\ \emph {et~al.}(2017)\citenamefont {Cross},
  \citenamefont {Bishop}, \citenamefont {Smolin},\ and\ \citenamefont
  {Gambetta}}]{cross2017open}%
  \BibitemOpen
  \bibfield  {author} {\bibinfo {author} {\bibfnamefont {A.~W.}\ \bibnamefont
  {Cross}}, \bibinfo {author} {\bibfnamefont {L.~S.}\ \bibnamefont {Bishop}},
  \bibinfo {author} {\bibfnamefont {J.~A.}\ \bibnamefont {Smolin}},\ and\
  \bibinfo {author} {\bibfnamefont {J.~M.}\ \bibnamefont {Gambetta}},\
  }\href@noop {} {\bibinfo {title} {Open quantum assembly language}} (\bibinfo
  {year} {2017}),\ \Eprint {https://arxiv.org/abs/1707.03429} {arXiv:1707.03429
  [quant-ph]} \BibitemShut {NoStop}%
\bibitem [{\citenamefont {McKay}\ \emph {et~al.}(2018)\citenamefont {McKay},
  \citenamefont {Alexander}, \citenamefont {Bello}, \citenamefont {Biercuk},
  \citenamefont {Bishop}, \citenamefont {Chen}, \citenamefont {Chow},
  \citenamefont {Córcoles}, \citenamefont {Egger}, \citenamefont {Filipp},
  \citenamefont {Gomez}, \citenamefont {Hush}, \citenamefont {Javadi-Abhari},
  \citenamefont {Moreda}, \citenamefont {Nation}, \citenamefont {Paulovicks},
  \citenamefont {Winston}, \citenamefont {Wood}, \citenamefont {Wootton},\ and\
  \citenamefont {Gambetta}}]{mckay2018qiskit}%
  \BibitemOpen
  \bibfield  {author} {\bibinfo {author} {\bibfnamefont {D.~C.}\ \bibnamefont
  {McKay}}, \bibinfo {author} {\bibfnamefont {T.}~\bibnamefont {Alexander}},
  \bibinfo {author} {\bibfnamefont {L.}~\bibnamefont {Bello}}, \bibinfo
  {author} {\bibfnamefont {M.~J.}\ \bibnamefont {Biercuk}}, \bibinfo {author}
  {\bibfnamefont {L.}~\bibnamefont {Bishop}}, \bibinfo {author} {\bibfnamefont
  {J.}~\bibnamefont {Chen}}, \bibinfo {author} {\bibfnamefont {J.~M.}\
  \bibnamefont {Chow}}, \bibinfo {author} {\bibfnamefont {A.~D.}\ \bibnamefont
  {Córcoles}}, \bibinfo {author} {\bibfnamefont {D.}~\bibnamefont {Egger}},
  \bibinfo {author} {\bibfnamefont {S.}~\bibnamefont {Filipp}}, \bibinfo
  {author} {\bibfnamefont {J.}~\bibnamefont {Gomez}}, \bibinfo {author}
  {\bibfnamefont {M.}~\bibnamefont {Hush}}, \bibinfo {author} {\bibfnamefont
  {A.}~\bibnamefont {Javadi-Abhari}}, \bibinfo {author} {\bibfnamefont
  {D.}~\bibnamefont {Moreda}}, \bibinfo {author} {\bibfnamefont
  {P.}~\bibnamefont {Nation}}, \bibinfo {author} {\bibfnamefont
  {B.}~\bibnamefont {Paulovicks}}, \bibinfo {author} {\bibfnamefont
  {E.}~\bibnamefont {Winston}}, \bibinfo {author} {\bibfnamefont {C.~J.}\
  \bibnamefont {Wood}}, \bibinfo {author} {\bibfnamefont {J.}~\bibnamefont
  {Wootton}},\ and\ \bibinfo {author} {\bibfnamefont {J.~M.}\ \bibnamefont
  {Gambetta}},\ }\href@noop {} {\bibinfo {title} {Qiskit backend specifications
  for openqasm and openpulse experiments}} (\bibinfo {year} {2018}),\ \Eprint
  {https://arxiv.org/abs/1809.03452} {arXiv:1809.03452 [quant-ph]} \BibitemShut
  {NoStop}%
\bibitem [{\citenamefont {Aleksandrowicz}\ \emph {et~al.}(2019)\citenamefont
  {Aleksandrowicz}, \citenamefont {Alexander}, \citenamefont {Barkoutsos},
  \citenamefont {Bello}, \citenamefont {Ben-Haim}, \citenamefont {Bucher},
  \citenamefont {Cabrera-Hern{\'a}ndez}, \citenamefont {Carballo-Franquis},
  \citenamefont {Chen}, \citenamefont {Chen} \emph
  {et~al.}}]{aleksandrowicz2019qiskit}%
  \BibitemOpen
  \bibfield  {author} {\bibinfo {author} {\bibfnamefont {G.}~\bibnamefont
  {Aleksandrowicz}}, \bibinfo {author} {\bibfnamefont {T.}~\bibnamefont
  {Alexander}}, \bibinfo {author} {\bibfnamefont {P.}~\bibnamefont
  {Barkoutsos}}, \bibinfo {author} {\bibfnamefont {L.}~\bibnamefont {Bello}},
  \bibinfo {author} {\bibfnamefont {Y.}~\bibnamefont {Ben-Haim}}, \bibinfo
  {author} {\bibfnamefont {D.}~\bibnamefont {Bucher}}, \bibinfo {author}
  {\bibfnamefont {F.}~\bibnamefont {Cabrera-Hern{\'a}ndez}}, \bibinfo {author}
  {\bibfnamefont {J.}~\bibnamefont {Carballo-Franquis}}, \bibinfo {author}
  {\bibfnamefont {A.}~\bibnamefont {Chen}}, \bibinfo {author} {\bibfnamefont
  {C.}~\bibnamefont {Chen}}, \emph {et~al.},\ }\href
  {https://doi.org/10.5281/zenodo.2562110} {\bibinfo {title} {Qiskit: An
  open-source framework for quantum computing}} (\bibinfo {year}
  {2019})\BibitemShut {NoStop}%
\bibitem [{\citenamefont {Khaneja}\ \emph {et~al.}(2005)\citenamefont
  {Khaneja}, \citenamefont {Reiss}, \citenamefont {Kehlet}, \citenamefont
  {Schulte-Herbr{\"u}ggen},\ and\ \citenamefont {Glaser}}]{khaneja2005optimal}%
  \BibitemOpen
  \bibfield  {author} {\bibinfo {author} {\bibfnamefont {N.}~\bibnamefont
  {Khaneja}}, \bibinfo {author} {\bibfnamefont {T.}~\bibnamefont {Reiss}},
  \bibinfo {author} {\bibfnamefont {C.}~\bibnamefont {Kehlet}}, \bibinfo
  {author} {\bibfnamefont {T.}~\bibnamefont {Schulte-Herbr{\"u}ggen}},\ and\
  \bibinfo {author} {\bibfnamefont {S.~J.}\ \bibnamefont {Glaser}},\ }\bibfield
   {title} {\bibinfo {title} {Optimal control of coupled spin dynamics: design
  of {NMR} pulse sequences by gradient ascent algorithms},\ }\href
  {https://doi.org/10.1016/j.jmr.2004.11.004} {\bibfield  {journal} {\bibinfo
  {journal} {Journal of magnetic resonance}\ }\textbf {\bibinfo {volume}
  {172}},\ \bibinfo {pages} {296} (\bibinfo {year} {2005})}\BibitemShut
  {NoStop}%
\bibitem [{\citenamefont {{de Fouquieres}}\ \emph {et~al.}(2011)\citenamefont
  {{de Fouquieres}}, \citenamefont {Schirmer}, \citenamefont {Glaser},\ and\
  \citenamefont {Kuprov}}]{de2011second}%
  \BibitemOpen
  \bibfield  {author} {\bibinfo {author} {\bibfnamefont {P.}~\bibnamefont {{de
  Fouquieres}}}, \bibinfo {author} {\bibfnamefont {S.}~\bibnamefont
  {Schirmer}}, \bibinfo {author} {\bibfnamefont {S.}~\bibnamefont {Glaser}},\
  and\ \bibinfo {author} {\bibfnamefont {I.}~\bibnamefont {Kuprov}},\
  }\bibfield  {title} {\bibinfo {title} {Second order gradient ascent pulse
  engineering},\ }\href {https://doi.org/10.1016/j.jmr.2011.07.023} {\bibfield
  {journal} {\bibinfo  {journal} {Journal of Magnetic Resonance}\ }\textbf
  {\bibinfo {volume} {212}},\ \bibinfo {pages} {412} (\bibinfo {year}
  {2011})}\BibitemShut {NoStop}%
\bibitem [{\citenamefont {Kelly}\ \emph {et~al.}(2014)\citenamefont {Kelly},
  \citenamefont {Barends}, \citenamefont {Campbell}, \citenamefont {Chen},
  \citenamefont {Chen}, \citenamefont {Chiaro}, \citenamefont {Dunsworth},
  \citenamefont {Fowler}, \citenamefont {Hoi}, \citenamefont {Jeffrey} \emph
  {et~al.}}]{kelly2014optimal}%
  \BibitemOpen
  \bibfield  {author} {\bibinfo {author} {\bibfnamefont {J.}~\bibnamefont
  {Kelly}}, \bibinfo {author} {\bibfnamefont {R.}~\bibnamefont {Barends}},
  \bibinfo {author} {\bibfnamefont {B.}~\bibnamefont {Campbell}}, \bibinfo
  {author} {\bibfnamefont {Y.}~\bibnamefont {Chen}}, \bibinfo {author}
  {\bibfnamefont {Z.}~\bibnamefont {Chen}}, \bibinfo {author} {\bibfnamefont
  {B.}~\bibnamefont {Chiaro}}, \bibinfo {author} {\bibfnamefont
  {A.}~\bibnamefont {Dunsworth}}, \bibinfo {author} {\bibfnamefont {A.~G.}\
  \bibnamefont {Fowler}}, \bibinfo {author} {\bibfnamefont {I.-C.}\
  \bibnamefont {Hoi}}, \bibinfo {author} {\bibfnamefont {E.}~\bibnamefont
  {Jeffrey}}, \emph {et~al.},\ }\bibfield  {title} {\bibinfo {title} {Optimal
  quantum control using randomized benchmarking},\ }\href
  {https://doi.org/10.1103/PhysRevLett.112.240504} {\bibfield  {journal}
  {\bibinfo  {journal} {Physical Review Letters}\ }\textbf {\bibinfo {volume}
  {112}},\ \bibinfo {pages} {240504} (\bibinfo {year} {2014})}\BibitemShut
  {NoStop}%
\bibitem [{\citenamefont {Glaser}\ \emph {et~al.}(2015)\citenamefont {Glaser},
  \citenamefont {Boscain}, \citenamefont {Calarco}, \citenamefont {Koch},
  \citenamefont {K{\"o}ckenberger}, \citenamefont {Kosloff}, \citenamefont
  {Kuprov}, \citenamefont {Luy}, \citenamefont {Schirmer}, \citenamefont
  {Schulte-Herbr{\"u}ggen} \emph {et~al.}}]{glaser2015training}%
  \BibitemOpen
  \bibfield  {author} {\bibinfo {author} {\bibfnamefont {S.~J.}\ \bibnamefont
  {Glaser}}, \bibinfo {author} {\bibfnamefont {U.}~\bibnamefont {Boscain}},
  \bibinfo {author} {\bibfnamefont {T.}~\bibnamefont {Calarco}}, \bibinfo
  {author} {\bibfnamefont {C.~P.}\ \bibnamefont {Koch}}, \bibinfo {author}
  {\bibfnamefont {W.}~\bibnamefont {K{\"o}ckenberger}}, \bibinfo {author}
  {\bibfnamefont {R.}~\bibnamefont {Kosloff}}, \bibinfo {author} {\bibfnamefont
  {I.}~\bibnamefont {Kuprov}}, \bibinfo {author} {\bibfnamefont
  {B.}~\bibnamefont {Luy}}, \bibinfo {author} {\bibfnamefont {S.}~\bibnamefont
  {Schirmer}}, \bibinfo {author} {\bibfnamefont {T.}~\bibnamefont
  {Schulte-Herbr{\"u}ggen}}, \emph {et~al.},\ }\bibfield  {title} {\bibinfo
  {title} {Training {S}chr{\"o}dinger’s cat: quantum optimal control},\
  }\href {https://doi.org/10.1140/epjd/e2015-60464-1} {\bibfield  {journal}
  {\bibinfo  {journal} {The European Physical Journal D}\ }\textbf {\bibinfo
  {volume} {69}},\ \bibinfo {pages} {1} (\bibinfo {year} {2015})}\BibitemShut
  {NoStop}%
\bibitem [{\citenamefont {Lu}\ \emph {et~al.}(2017)\citenamefont {Lu},
  \citenamefont {Li}, \citenamefont {Li}, \citenamefont {Katiyar},
  \citenamefont {Park}, \citenamefont {Feng}, \citenamefont {Xin},
  \citenamefont {Li}, \citenamefont {Long}, \citenamefont {Brodutch} \emph
  {et~al.}}]{lu2017enhancing}%
  \BibitemOpen
  \bibfield  {author} {\bibinfo {author} {\bibfnamefont {D.}~\bibnamefont
  {Lu}}, \bibinfo {author} {\bibfnamefont {K.}~\bibnamefont {Li}}, \bibinfo
  {author} {\bibfnamefont {J.}~\bibnamefont {Li}}, \bibinfo {author}
  {\bibfnamefont {H.}~\bibnamefont {Katiyar}}, \bibinfo {author} {\bibfnamefont
  {A.~J.}\ \bibnamefont {Park}}, \bibinfo {author} {\bibfnamefont
  {G.}~\bibnamefont {Feng}}, \bibinfo {author} {\bibfnamefont {T.}~\bibnamefont
  {Xin}}, \bibinfo {author} {\bibfnamefont {H.}~\bibnamefont {Li}}, \bibinfo
  {author} {\bibfnamefont {G.}~\bibnamefont {Long}}, \bibinfo {author}
  {\bibfnamefont {A.}~\bibnamefont {Brodutch}}, \emph {et~al.},\ }\bibfield
  {title} {\bibinfo {title} {Enhancing quantum control by bootstrapping a
  quantum processor of 12 qubits},\ }\href
  {https://doi.org/10.1038/s41534-017-0045-z} {\bibfield  {journal} {\bibinfo
  {journal} {npj Quantum Information}\ }\textbf {\bibinfo {volume} {3}},\
  \bibinfo {pages} {1} (\bibinfo {year} {2017})}\BibitemShut {NoStop}%
\bibitem [{\citenamefont {Gokhale}\ \emph {et~al.}(2019)\citenamefont
  {Gokhale}, \citenamefont {Ding}, \citenamefont {Propson}, \citenamefont
  {Winkler}, \citenamefont {Leung}, \citenamefont {Shi}, \citenamefont
  {Schuster}, \citenamefont {Hoffmann},\ and\ \citenamefont
  {Chong}}]{gokhale2019partial}%
  \BibitemOpen
  \bibfield  {author} {\bibinfo {author} {\bibfnamefont {P.}~\bibnamefont
  {Gokhale}}, \bibinfo {author} {\bibfnamefont {Y.}~\bibnamefont {Ding}},
  \bibinfo {author} {\bibfnamefont {T.}~\bibnamefont {Propson}}, \bibinfo
  {author} {\bibfnamefont {C.}~\bibnamefont {Winkler}}, \bibinfo {author}
  {\bibfnamefont {N.}~\bibnamefont {Leung}}, \bibinfo {author} {\bibfnamefont
  {Y.}~\bibnamefont {Shi}}, \bibinfo {author} {\bibfnamefont {D.~I.}\
  \bibnamefont {Schuster}}, \bibinfo {author} {\bibfnamefont {H.}~\bibnamefont
  {Hoffmann}},\ and\ \bibinfo {author} {\bibfnamefont {F.~T.}\ \bibnamefont
  {Chong}},\ }\bibfield  {title} {\bibinfo {title} {Partial compilation of
  variational algorithms for noisy intermediate-scale quantum machines},\ }in\
  \href {https://doi.org/10.1145/3352460.3358313} {\emph {\bibinfo {booktitle}
  {Proceedings of the 52nd Annual IEEE/ACM International Symposium on
  Microarchitecture}}}\ (\bibinfo {year} {2019})\ pp.\ \bibinfo {pages}
  {266--278}\BibitemShut {NoStop}%
\bibitem [{\citenamefont {{Chen}}\ \emph {et~al.}(2014)\citenamefont {{Chen}},
  \citenamefont {{Dong}}, \citenamefont {{Li}}, \citenamefont {{Chu}},\ and\
  \citenamefont {{Tarn}}}]{chen2014fidelity}%
  \BibitemOpen
  \bibfield  {author} {\bibinfo {author} {\bibfnamefont {C.}~\bibnamefont
  {{Chen}}}, \bibinfo {author} {\bibfnamefont {D.}~\bibnamefont {{Dong}}},
  \bibinfo {author} {\bibfnamefont {H.}~\bibnamefont {{Li}}}, \bibinfo {author}
  {\bibfnamefont {J.}~\bibnamefont {{Chu}}},\ and\ \bibinfo {author}
  {\bibfnamefont {T.}~\bibnamefont {{Tarn}}},\ }\bibfield  {title} {\bibinfo
  {title} {Fidelity-based probabilistic {Q}-learning for control of quantum
  systems},\ }\href {https://doi.org/10.1109/TNNLS.2013.2283574} {\bibfield
  {journal} {\bibinfo  {journal} {IEEE Transactions on Neural Networks and
  Learning Systems}\ }\textbf {\bibinfo {volume} {25}},\ \bibinfo {pages} {920}
  (\bibinfo {year} {2014})}\BibitemShut {NoStop}%
\bibitem [{\citenamefont {Bukov}\ \emph {et~al.}(2018)\citenamefont {Bukov},
  \citenamefont {Day}, \citenamefont {Sels}, \citenamefont {Weinberg},
  \citenamefont {Polkovnikov},\ and\ \citenamefont
  {Mehta}}]{bukov2018reinforcement}%
  \BibitemOpen
  \bibfield  {author} {\bibinfo {author} {\bibfnamefont {M.}~\bibnamefont
  {Bukov}}, \bibinfo {author} {\bibfnamefont {A.~G.}\ \bibnamefont {Day}},
  \bibinfo {author} {\bibfnamefont {D.}~\bibnamefont {Sels}}, \bibinfo {author}
  {\bibfnamefont {P.}~\bibnamefont {Weinberg}}, \bibinfo {author}
  {\bibfnamefont {A.}~\bibnamefont {Polkovnikov}},\ and\ \bibinfo {author}
  {\bibfnamefont {P.}~\bibnamefont {Mehta}},\ }\bibfield  {title} {\bibinfo
  {title} {Reinforcement learning in different phases of quantum control},\
  }\href {https://doi.org/10.1103/PhysRevX.8.031086} {\bibfield  {journal}
  {\bibinfo  {journal} {Physical Review X}\ }\textbf {\bibinfo {volume} {8}},\
  \bibinfo {pages} {031086} (\bibinfo {year} {2018})}\BibitemShut {NoStop}%
\bibitem [{\citenamefont {Zhang}\ \emph {et~al.}(2019)\citenamefont {Zhang},
  \citenamefont {Wei}, \citenamefont {Asad}, \citenamefont {Yang},\ and\
  \citenamefont {Wang}}]{zhang2019does}%
  \BibitemOpen
  \bibfield  {author} {\bibinfo {author} {\bibfnamefont {X.-M.}\ \bibnamefont
  {Zhang}}, \bibinfo {author} {\bibfnamefont {Z.}~\bibnamefont {Wei}}, \bibinfo
  {author} {\bibfnamefont {R.}~\bibnamefont {Asad}}, \bibinfo {author}
  {\bibfnamefont {X.-C.}\ \bibnamefont {Yang}},\ and\ \bibinfo {author}
  {\bibfnamefont {X.}~\bibnamefont {Wang}},\ }\bibfield  {title} {\bibinfo
  {title} {When does reinforcement learning stand out in quantum control? a
  comparative study on state preparation},\ }\href
  {https://doi.org/10.1038/s41534-019-0201-8} {\bibfield  {journal} {\bibinfo
  {journal} {npj Quantum Information}\ }\textbf {\bibinfo {volume} {5}},\
  \bibinfo {pages} {1} (\bibinfo {year} {2019})}\BibitemShut {NoStop}%
\bibitem [{\citenamefont {Niu}\ \emph {et~al.}(2019)\citenamefont {Niu},
  \citenamefont {Boixo}, \citenamefont {Smelyanskiy},\ and\ \citenamefont
  {Neven}}]{niu2019universal}%
  \BibitemOpen
  \bibfield  {author} {\bibinfo {author} {\bibfnamefont {M.~Y.}\ \bibnamefont
  {Niu}}, \bibinfo {author} {\bibfnamefont {S.}~\bibnamefont {Boixo}}, \bibinfo
  {author} {\bibfnamefont {V.~N.}\ \bibnamefont {Smelyanskiy}},\ and\ \bibinfo
  {author} {\bibfnamefont {H.}~\bibnamefont {Neven}},\ }\bibfield  {title}
  {\bibinfo {title} {Universal quantum control through deep reinforcement
  learning},\ }\href {https://doi.org/10.1038/s41534-019-0141-3} {\bibfield
  {journal} {\bibinfo  {journal} {npj Quantum Information}\ }\textbf {\bibinfo
  {volume} {5}},\ \bibinfo {pages} {1} (\bibinfo {year} {2019})}\BibitemShut
  {NoStop}%
\bibitem [{\citenamefont {Stenger}\ \emph {et~al.}(2020)\citenamefont
  {Stenger}, \citenamefont {Bronn}, \citenamefont {Egger},\ and\ \citenamefont
  {Pekker}}]{stenger2020simulating}%
  \BibitemOpen
  \bibfield  {author} {\bibinfo {author} {\bibfnamefont {J.~P.~T.}\
  \bibnamefont {Stenger}}, \bibinfo {author} {\bibfnamefont {N.~T.}\
  \bibnamefont {Bronn}}, \bibinfo {author} {\bibfnamefont {D.~J.}\ \bibnamefont
  {Egger}},\ and\ \bibinfo {author} {\bibfnamefont {D.}~\bibnamefont
  {Pekker}},\ }\href@noop {} {\bibinfo {title} {Simulating the dynamics of
  braiding of majorana zero modes using an ibm quantum computer}} (\bibinfo
  {year} {2020}),\ \Eprint {https://arxiv.org/abs/2012.11660} {arXiv:2012.11660
  [quant-ph]} \BibitemShut {NoStop}%
\bibitem [{\citenamefont {Alexander}\ \emph {et~al.}(2020)\citenamefont
  {Alexander}, \citenamefont {Kanazawa}, \citenamefont {Egger}, \citenamefont
  {Capelluto}, \citenamefont {Wood}, \citenamefont {Javadi-Abhari},\ and\
  \citenamefont {C~McKay}}]{alexander2020qiskit}%
  \BibitemOpen
  \bibfield  {author} {\bibinfo {author} {\bibfnamefont {T.}~\bibnamefont
  {Alexander}}, \bibinfo {author} {\bibfnamefont {N.}~\bibnamefont {Kanazawa}},
  \bibinfo {author} {\bibfnamefont {D.~J.}\ \bibnamefont {Egger}}, \bibinfo
  {author} {\bibfnamefont {L.}~\bibnamefont {Capelluto}}, \bibinfo {author}
  {\bibfnamefont {C.~J.}\ \bibnamefont {Wood}}, \bibinfo {author}
  {\bibfnamefont {A.}~\bibnamefont {Javadi-Abhari}},\ and\ \bibinfo {author}
  {\bibfnamefont {D.}~\bibnamefont {C~McKay}},\ }\bibfield  {title} {\bibinfo
  {title} {Qiskit pulse: programming quantum computers through the cloud with
  pulses},\ }\href {https://doi.org/10.1088/2058-9565/aba404} {\bibfield
  {journal} {\bibinfo  {journal} {Quantum Science and Technology}\ }\textbf
  {\bibinfo {volume} {5}},\ \bibinfo {pages} {044006} (\bibinfo {year}
  {2020})}\BibitemShut {NoStop}%
\bibitem [{\citenamefont {Garion}\ \emph {et~al.}(2020)\citenamefont {Garion},
  \citenamefont {Kanazawa}, \citenamefont {Landa}, \citenamefont {McKay},
  \citenamefont {Sheldon}, \citenamefont {Cross},\ and\ \citenamefont
  {Wood}}]{garion2020experimental}%
  \BibitemOpen
  \bibfield  {author} {\bibinfo {author} {\bibfnamefont {S.}~\bibnamefont
  {Garion}}, \bibinfo {author} {\bibfnamefont {N.}~\bibnamefont {Kanazawa}},
  \bibinfo {author} {\bibfnamefont {H.}~\bibnamefont {Landa}}, \bibinfo
  {author} {\bibfnamefont {D.~C.}\ \bibnamefont {McKay}}, \bibinfo {author}
  {\bibfnamefont {S.}~\bibnamefont {Sheldon}}, \bibinfo {author} {\bibfnamefont
  {A.~W.}\ \bibnamefont {Cross}},\ and\ \bibinfo {author} {\bibfnamefont
  {C.~J.}\ \bibnamefont {Wood}},\ }\href@noop {} {\bibinfo {title}
  {Experimental implementation of non-clifford interleaved randomized
  benchmarking with a controlled-s gate}} (\bibinfo {year} {2020}),\ \Eprint
  {https://arxiv.org/abs/2007.08532} {arXiv:2007.08532 [quant-ph]} \BibitemShut
  {NoStop}%
\bibitem [{\citenamefont {Bastidas}\ \emph {et~al.}(2020)\citenamefont
  {Bastidas}, \citenamefont {Haug}, \citenamefont {Gravel}, \citenamefont
  {Kwek}, \citenamefont {Munro},\ and\ \citenamefont
  {Nemoto}}]{bastidas2020fullyprogrammable}%
  \BibitemOpen
  \bibfield  {author} {\bibinfo {author} {\bibfnamefont {V.~M.}\ \bibnamefont
  {Bastidas}}, \bibinfo {author} {\bibfnamefont {T.}~\bibnamefont {Haug}},
  \bibinfo {author} {\bibfnamefont {C.}~\bibnamefont {Gravel}}, \bibinfo
  {author} {\bibfnamefont {L.~C.}\ \bibnamefont {Kwek}}, \bibinfo {author}
  {\bibfnamefont {W.~J.}\ \bibnamefont {Munro}},\ and\ \bibinfo {author}
  {\bibfnamefont {K.}~\bibnamefont {Nemoto}},\ }\href@noop {} {\bibinfo {title}
  {Fully-programmable universal quantum simulator with a one-dimensional
  quantum processor}} (\bibinfo {year} {2020}),\ \Eprint
  {https://arxiv.org/abs/2009.00823} {arXiv:2009.00823 [quant-ph]} \BibitemShut
  {NoStop}%
\bibitem [{\citenamefont {Malz}\ and\ \citenamefont
  {Smith}(2020)}]{malz2020topological}%
  \BibitemOpen
  \bibfield  {author} {\bibinfo {author} {\bibfnamefont {D.}~\bibnamefont
  {Malz}}\ and\ \bibinfo {author} {\bibfnamefont {A.}~\bibnamefont {Smith}},\
  }\href@noop {} {\bibinfo {title} {Topological two-dimensional floquet lattice
  on a single superconducting qubit}} (\bibinfo {year} {2020}),\ \Eprint
  {https://arxiv.org/abs/2012.01459} {arXiv:2012.01459 [quant-ph]} \BibitemShut
  {NoStop}%
\bibitem [{\citenamefont {Floquet}(1883)}]{floquet1883equations}%
  \BibitemOpen
  \bibfield  {author} {\bibinfo {author} {\bibfnamefont {G.}~\bibnamefont
  {Floquet}},\ }\bibfield  {title} {\bibinfo {title} {Sur les {\'e}quations
  diff{\'e}rentielles lin{\'e}aires {\`a} coefficients p{\'e}riodiques},\ }in\
  \href {https://doi.org/10.24033/asens.220} {\emph {\bibinfo {booktitle}
  {Annales scientifiques de l'{\'E}cole normale sup{\'e}rieure}}},\
  Vol.~\bibinfo {volume} {12}\ (\bibinfo {year} {1883})\ pp.\ \bibinfo {pages}
  {47--88}\BibitemShut {NoStop}%
\bibitem [{\citenamefont {Shirley}(1965)}]{shirley1965solution}%
  \BibitemOpen
  \bibfield  {author} {\bibinfo {author} {\bibfnamefont {J.~H.}\ \bibnamefont
  {Shirley}},\ }\bibfield  {title} {\bibinfo {title} {Solution of the
  {Schr{\"o}dinger} equation with a {Hamiltonian} periodic in time},\ }\href
  {https://doi.org/10.1103/PhysRev.138.B979} {\bibfield  {journal} {\bibinfo
  {journal} {Physical Review}\ }\textbf {\bibinfo {volume} {138}},\ \bibinfo
  {pages} {B979} (\bibinfo {year} {1965})}\BibitemShut {NoStop}%
\bibitem [{\citenamefont {Russomanno}\ and\ \citenamefont
  {Santoro}(2017)}]{russomanno2017floquet}%
  \BibitemOpen
  \bibfield  {author} {\bibinfo {author} {\bibfnamefont {A.}~\bibnamefont
  {Russomanno}}\ and\ \bibinfo {author} {\bibfnamefont {G.~E.}\ \bibnamefont
  {Santoro}},\ }\bibfield  {title} {\bibinfo {title} {Floquet resonances close
  to the adiabatic limit and the effect of dissipation},\ }\href
  {https://doi.org/10.1088/1742-5468/aa8702} {\bibfield  {journal} {\bibinfo
  {journal} {Journal of Statistical Mechanics: Theory and Experiment}\ }\textbf
  {\bibinfo {volume} {2017}},\ \bibinfo {pages} {103104} (\bibinfo {year}
  {2017})}\BibitemShut {NoStop}%
\bibitem [{\citenamefont {Gagge}\ and\ \citenamefont
  {Larson}(2018)}]{gagge2018bloch}%
  \BibitemOpen
  \bibfield  {author} {\bibinfo {author} {\bibfnamefont {A.}~\bibnamefont
  {Gagge}}\ and\ \bibinfo {author} {\bibfnamefont {J.}~\bibnamefont {Larson}},\
  }\bibfield  {title} {\bibinfo {title} {Bloch-like energy oscillations},\
  }\href {https://doi.org/10.1103/PhysRevA.98.053820} {\bibfield  {journal}
  {\bibinfo  {journal} {Physical Review A}\ }\textbf {\bibinfo {volume} {98}},\
  \bibinfo {pages} {053820} (\bibinfo {year} {2018})}\BibitemShut {NoStop}%
\bibitem [{Note2()}]{Note2}%
  \BibitemOpen
  \bibinfo {note} {Remarkably, this approximation remains valid even in
  situations where the system shows a dynamical instability and the energy
  absorption rate diverges. For example, the parametric resonance of an
  harmonic oscillator can be described by considering only two Fourier
  harmonics \cite {landau1976mechanics}. At the resonance, the Floquet
  functions diverge in Fock space of the oscillator, but remain localized in
  Floquet space \cite {weigert2002quantum}.}\BibitemShut {Stop}%
\bibitem [{\citenamefont {Chirikov}\ and\ \citenamefont
  {Shepelyansky}(2008)}]{chirikov2008chirikov}%
  \BibitemOpen
  \bibfield  {author} {\bibinfo {author} {\bibfnamefont {B.}~\bibnamefont
  {Chirikov}}\ and\ \bibinfo {author} {\bibfnamefont {D.}~\bibnamefont
  {Shepelyansky}},\ }\bibfield  {title} {\bibinfo {title} {Chirikov standard
  map},\ }\href {https://doi.org/10.4249/scholarpedia.3550} {\bibfield
  {journal} {\bibinfo  {journal} {Scholarpedia}\ }\textbf {\bibinfo {volume}
  {3}},\ \bibinfo {pages} {3550} (\bibinfo {year} {2008})}\BibitemShut
  {NoStop}%
\bibitem [{Note3()}]{Note3}%
  \BibitemOpen
  \bibinfo {note} {The delocalization of the Floquet eigenfunctions is a
  necessary condition for the dynamical localization in quantum models of
  kicked rotors \cite {grempel1982localization,fishman1982chaos}.}\BibitemShut
  {Stop}%
\bibitem [{Note4()}]{Note4}%
  \BibitemOpen
  \bibinfo {note} {The Hamiltonian (\ref {eq:Ht}) is actually realized in a
  rotating frame, such that $\omega _z$ is the detuning between the driving
  field and the bare frequency of the qubit, and $h_x(t)$ is one quadrature of
  the driving field. We point out that the other quadrature enables one to
  realize lattice model with imaginary hopping terms.}\BibitemShut {Stop}%
\bibitem [{\citenamefont {Bellman}\ \emph {et~al.}(1956)\citenamefont
  {Bellman}, \citenamefont {Glicksberg},\ and\ \citenamefont
  {Gross}}]{bellman1956bang}%
  \BibitemOpen
  \bibfield  {author} {\bibinfo {author} {\bibfnamefont {R.}~\bibnamefont
  {Bellman}}, \bibinfo {author} {\bibfnamefont {I.}~\bibnamefont
  {Glicksberg}},\ and\ \bibinfo {author} {\bibfnamefont {O.}~\bibnamefont
  {Gross}},\ }\bibfield  {title} {\bibinfo {title} {On the “bang-bang”
  control problem},\ }\href {https://doi.org/10.1090/qam/78516} {\bibfield
  {journal} {\bibinfo  {journal} {Quarterly of Applied Mathematics}\ }\textbf
  {\bibinfo {volume} {14}},\ \bibinfo {pages} {11} (\bibinfo {year}
  {1956})}\BibitemShut {NoStop}%
\bibitem [{\citenamefont {Viola}\ and\ \citenamefont
  {Lloyd}(1998)}]{viola1998dynamical}%
  \BibitemOpen
  \bibfield  {author} {\bibinfo {author} {\bibfnamefont {L.}~\bibnamefont
  {Viola}}\ and\ \bibinfo {author} {\bibfnamefont {S.}~\bibnamefont {Lloyd}},\
  }\bibfield  {title} {\bibinfo {title} {Dynamical suppression of decoherence
  in two-state quantum systems},\ }\href
  {https://doi.org/10.1103/PhysRevA.58.2733} {\bibfield  {journal} {\bibinfo
  {journal} {Physical Review A}\ }\textbf {\bibinfo {volume} {58}},\ \bibinfo
  {pages} {2733} (\bibinfo {year} {1998})}\BibitemShut {NoStop}%
\bibitem [{ibm(2021)}]{ibm2021}%
  \BibitemOpen
  \href@noop {} {\bibinfo {title} {{IBM} {Q}uantum}},\ \bibinfo {howpublished}
  {\url{{https://quantum-computing.ibm.com}}} (\bibinfo {year} {2021}),\
  \bibinfo {note} {2021}\BibitemShut {NoStop}%
\bibitem [{Note5()}]{Note5}%
  \BibitemOpen
  \bibinfo {note} {Pauli matrices are examples of bounded operators with norm
  $\norm *{\protect \hat {O}}=1$.}\BibitemShut {Stop}%
\bibitem [{\citenamefont {Bassman}\ \emph {et~al.}(2021)\citenamefont
  {Bassman}, \citenamefont {Urbanek}, \citenamefont {Metcalf}, \citenamefont
  {Carter}, \citenamefont {Kemper},\ and\ \citenamefont
  {de~Jong}}]{bassman2021simulating}%
  \BibitemOpen
  \bibfield  {author} {\bibinfo {author} {\bibfnamefont {L.}~\bibnamefont
  {Bassman}}, \bibinfo {author} {\bibfnamefont {M.}~\bibnamefont {Urbanek}},
  \bibinfo {author} {\bibfnamefont {M.}~\bibnamefont {Metcalf}}, \bibinfo
  {author} {\bibfnamefont {J.}~\bibnamefont {Carter}}, \bibinfo {author}
  {\bibfnamefont {A.~F.}\ \bibnamefont {Kemper}},\ and\ \bibinfo {author}
  {\bibfnamefont {W.}~\bibnamefont {de~Jong}},\ }\href@noop {} {\bibinfo
  {title} {Simulating quantum materials with digital quantum computers}}
  (\bibinfo {year} {2021}),\ \Eprint {https://arxiv.org/abs/2101.08836}
  {arXiv:2101.08836 [quant-ph]} \BibitemShut {NoStop}%
\bibitem [{\citenamefont {Bharti}\ \emph {et~al.}(2021)\citenamefont {Bharti},
  \citenamefont {Cervera-Lierta}, \citenamefont {Kyaw}, \citenamefont {Haug},
  \citenamefont {Alperin-Lea}, \citenamefont {Anand}, \citenamefont {Degroote},
  \citenamefont {Heimonen}, \citenamefont {Kottmann}, \citenamefont {Menke},
  \citenamefont {Mok}, \citenamefont {Sim}, \citenamefont {Kwek},\ and\
  \citenamefont {Aspuru-Guzik}}]{bharti2021noisy}%
  \BibitemOpen
  \bibfield  {author} {\bibinfo {author} {\bibfnamefont {K.}~\bibnamefont
  {Bharti}}, \bibinfo {author} {\bibfnamefont {A.}~\bibnamefont
  {Cervera-Lierta}}, \bibinfo {author} {\bibfnamefont {T.~H.}\ \bibnamefont
  {Kyaw}}, \bibinfo {author} {\bibfnamefont {T.}~\bibnamefont {Haug}}, \bibinfo
  {author} {\bibfnamefont {S.}~\bibnamefont {Alperin-Lea}}, \bibinfo {author}
  {\bibfnamefont {A.}~\bibnamefont {Anand}}, \bibinfo {author} {\bibfnamefont
  {M.}~\bibnamefont {Degroote}}, \bibinfo {author} {\bibfnamefont
  {H.}~\bibnamefont {Heimonen}}, \bibinfo {author} {\bibfnamefont {J.~S.}\
  \bibnamefont {Kottmann}}, \bibinfo {author} {\bibfnamefont {T.}~\bibnamefont
  {Menke}}, \bibinfo {author} {\bibfnamefont {W.-K.}\ \bibnamefont {Mok}},
  \bibinfo {author} {\bibfnamefont {S.}~\bibnamefont {Sim}}, \bibinfo {author}
  {\bibfnamefont {L.-C.}\ \bibnamefont {Kwek}},\ and\ \bibinfo {author}
  {\bibfnamefont {A.}~\bibnamefont {Aspuru-Guzik}},\ }\href@noop {} {\bibinfo
  {title} {Noisy intermediate-scale quantum (nisq) algorithms}} (\bibinfo
  {year} {2021}),\ \Eprint {https://arxiv.org/abs/2101.08448} {arXiv:2101.08448
  [quant-ph]} \BibitemShut {NoStop}%
\bibitem [{\citenamefont {Roushan}\ \emph {et~al.}(2017)\citenamefont
  {Roushan}, \citenamefont {Neill}, \citenamefont {Tangpanitanon},
  \citenamefont {Bastidas}, \citenamefont {Megrant}, \citenamefont {Barends},
  \citenamefont {Chen}, \citenamefont {Chen}, \citenamefont {Chiaro},
  \citenamefont {Dunsworth}, \citenamefont {Fowler}, \citenamefont {Foxen},
  \citenamefont {Giustina}, \citenamefont {Jeffrey}, \citenamefont {Kelly},
  \citenamefont {Lucero}, \citenamefont {Mutus}, \citenamefont {Neeley},
  \citenamefont {Quintana}, \citenamefont {Sank}, \citenamefont {Vainsencher},
  \citenamefont {Wenner}, \citenamefont {White}, \citenamefont {Neven},
  \citenamefont {Angelakis},\ and\ \citenamefont
  {Martinis}}]{roushan2017spectroscopic}%
  \BibitemOpen
  \bibfield  {author} {\bibinfo {author} {\bibfnamefont {P.}~\bibnamefont
  {Roushan}}, \bibinfo {author} {\bibfnamefont {C.}~\bibnamefont {Neill}},
  \bibinfo {author} {\bibfnamefont {J.}~\bibnamefont {Tangpanitanon}}, \bibinfo
  {author} {\bibfnamefont {V.~M.}\ \bibnamefont {Bastidas}}, \bibinfo {author}
  {\bibfnamefont {A.}~\bibnamefont {Megrant}}, \bibinfo {author} {\bibfnamefont
  {R.}~\bibnamefont {Barends}}, \bibinfo {author} {\bibfnamefont
  {Y.}~\bibnamefont {Chen}}, \bibinfo {author} {\bibfnamefont {Z.}~\bibnamefont
  {Chen}}, \bibinfo {author} {\bibfnamefont {B.}~\bibnamefont {Chiaro}},
  \bibinfo {author} {\bibfnamefont {A.}~\bibnamefont {Dunsworth}}, \bibinfo
  {author} {\bibfnamefont {A.}~\bibnamefont {Fowler}}, \bibinfo {author}
  {\bibfnamefont {B.}~\bibnamefont {Foxen}}, \bibinfo {author} {\bibfnamefont
  {M.}~\bibnamefont {Giustina}}, \bibinfo {author} {\bibfnamefont
  {E.}~\bibnamefont {Jeffrey}}, \bibinfo {author} {\bibfnamefont
  {J.}~\bibnamefont {Kelly}}, \bibinfo {author} {\bibfnamefont
  {E.}~\bibnamefont {Lucero}}, \bibinfo {author} {\bibfnamefont
  {J.}~\bibnamefont {Mutus}}, \bibinfo {author} {\bibfnamefont
  {M.}~\bibnamefont {Neeley}}, \bibinfo {author} {\bibfnamefont
  {C.}~\bibnamefont {Quintana}}, \bibinfo {author} {\bibfnamefont
  {D.}~\bibnamefont {Sank}}, \bibinfo {author} {\bibfnamefont {A.}~\bibnamefont
  {Vainsencher}}, \bibinfo {author} {\bibfnamefont {J.}~\bibnamefont {Wenner}},
  \bibinfo {author} {\bibfnamefont {T.}~\bibnamefont {White}}, \bibinfo
  {author} {\bibfnamefont {H.}~\bibnamefont {Neven}}, \bibinfo {author}
  {\bibfnamefont {D.~G.}\ \bibnamefont {Angelakis}},\ and\ \bibinfo {author}
  {\bibfnamefont {J.}~\bibnamefont {Martinis}},\ }\bibfield  {title} {\bibinfo
  {title} {Spectroscopic signatures of localization with interacting photons in
  superconducting qubits},\ }\href {https://doi.org/10.1126/science.aao1401}
  {\bibfield  {journal} {\bibinfo  {journal} {Science}\ }\textbf {\bibinfo
  {volume} {358}},\ \bibinfo {pages} {1175} (\bibinfo {year}
  {2017})}\BibitemShut {NoStop}%
\bibitem [{\citenamefont {Smith}\ \emph {et~al.}(2019)\citenamefont {Smith},
  \citenamefont {Kim}, \citenamefont {Pollmann},\ and\ \citenamefont
  {Knolle}}]{smith2019simulating}%
  \BibitemOpen
  \bibfield  {author} {\bibinfo {author} {\bibfnamefont {A.}~\bibnamefont
  {Smith}}, \bibinfo {author} {\bibfnamefont {M.}~\bibnamefont {Kim}}, \bibinfo
  {author} {\bibfnamefont {F.}~\bibnamefont {Pollmann}},\ and\ \bibinfo
  {author} {\bibfnamefont {J.}~\bibnamefont {Knolle}},\ }\bibfield  {title}
  {\bibinfo {title} {Simulating quantum many-body dynamics on a current digital
  quantum computer},\ }\href {https://doi.org/10.1038/s41534-019-0217-0}
  {\bibfield  {journal} {\bibinfo  {journal} {npj Quantum Information}\
  }\textbf {\bibinfo {volume} {5}},\ \bibinfo {pages} {1} (\bibinfo {year}
  {2019})}\BibitemShut {NoStop}%
\bibitem [{\citenamefont {Mei}\ \emph {et~al.}(2020)\citenamefont {Mei},
  \citenamefont {Guo}, \citenamefont {Yu}, \citenamefont {Xiao}, \citenamefont
  {Zhu},\ and\ \citenamefont {Jia}}]{mei2020digital}%
  \BibitemOpen
  \bibfield  {author} {\bibinfo {author} {\bibfnamefont {F.}~\bibnamefont
  {Mei}}, \bibinfo {author} {\bibfnamefont {Q.}~\bibnamefont {Guo}}, \bibinfo
  {author} {\bibfnamefont {Y.-F.}\ \bibnamefont {Yu}}, \bibinfo {author}
  {\bibfnamefont {L.}~\bibnamefont {Xiao}}, \bibinfo {author} {\bibfnamefont
  {S.-L.}\ \bibnamefont {Zhu}},\ and\ \bibinfo {author} {\bibfnamefont
  {S.}~\bibnamefont {Jia}},\ }\bibfield  {title} {\bibinfo {title} {Digital
  simulation of topological matter on programmable quantum processors},\ }\href
  {https://doi.org/10.1103/PhysRevLett.125.160503} {\bibfield  {journal}
  {\bibinfo  {journal} {Physical Review Letters}\ }\textbf {\bibinfo {volume}
  {125}},\ \bibinfo {pages} {160503} (\bibinfo {year} {2020})}\BibitemShut
  {NoStop}%
\bibitem [{\citenamefont {Roses}\ \emph {et~al.}(2021)\citenamefont {Roses},
  \citenamefont {Landa},\ and\ \citenamefont
  {Dalla~Torre}}]{roses2021simulating}%
  \BibitemOpen
  \bibfield  {author} {\bibinfo {author} {\bibfnamefont {M.~M.}\ \bibnamefont
  {Roses}}, \bibinfo {author} {\bibfnamefont {H.}~\bibnamefont {Landa}},\ and\
  \bibinfo {author} {\bibfnamefont {E.~G.}\ \bibnamefont {Dalla~Torre}},\
  }\href {https://doi.org/10.5281/zenodo.4549200} {\bibinfo {title} {Simulating
  long-range hopping with periodically-driven superconducting qubits --- data
  files}} (\bibinfo {year} {2021})\BibitemShut {NoStop}%
\bibitem [{Note6()}]{Note6}%
  \BibitemOpen
  \bibinfo {note} {For the sake of brevity we limit our operators to those that
  have a norm of 1, \protect \emph {i.e.\spacefactor \@m {}}, $\norm *{\protect
  \hat {O}}=1$}\BibitemShut {NoStop}%
\bibitem [{\citenamefont {Rudner}\ and\ \citenamefont
  {Lindner}(2020)}]{rudner2020floquet}%
  \BibitemOpen
  \bibfield  {author} {\bibinfo {author} {\bibfnamefont {M.~S.}\ \bibnamefont
  {Rudner}}\ and\ \bibinfo {author} {\bibfnamefont {N.~H.}\ \bibnamefont
  {Lindner}},\ }\href@noop {} {\bibinfo {title} {The floquet engineer's
  handbook}} (\bibinfo {year} {2020}),\ \Eprint
  {https://arxiv.org/abs/2003.08252} {arXiv:2003.08252 [cond-mat.mes-hall]}
  \BibitemShut {NoStop}%
\bibitem [{\citenamefont {Landau}\ and\ \citenamefont
  {Lifshitz}(1976)}]{landau1976mechanics}%
  \BibitemOpen
  \bibfield  {author} {\bibinfo {author} {\bibfnamefont {L.~D.}\ \bibnamefont
  {Landau}}\ and\ \bibinfo {author} {\bibfnamefont {E.~M.}\ \bibnamefont
  {Lifshitz}},\ }\href {https://books.google.co.il/books?id=e-xASAehg1sC}
  {\emph {\bibinfo {title} {Mechanics: Volume 1}}},\ Vol.~\bibinfo {volume}
  {1}\ (\bibinfo  {publisher} {Butterworth-Heinemann},\ \bibinfo {year}
  {1976})\BibitemShut {NoStop}%
\bibitem [{\citenamefont {Weigert}(2002)}]{weigert2002quantum}%
  \BibitemOpen
  \bibfield  {author} {\bibinfo {author} {\bibfnamefont {S.}~\bibnamefont
  {Weigert}},\ }\bibfield  {title} {\bibinfo {title} {Quantum parametric
  resonance},\ }\href@noop {} {\bibfield  {journal} {\bibinfo  {journal}
  {Journal of Physics A: Mathematical and General}\ }\textbf {\bibinfo {volume}
  {35}},\ \bibinfo {pages} {4169} (\bibinfo {year} {2002})}\BibitemShut
  {NoStop}%
\bibitem [{\citenamefont {Grempel}\ \emph {et~al.}(1982)\citenamefont
  {Grempel}, \citenamefont {Fishman},\ and\ \citenamefont
  {Prange}}]{grempel1982localization}%
  \BibitemOpen
  \bibfield  {author} {\bibinfo {author} {\bibfnamefont {D.}~\bibnamefont
  {Grempel}}, \bibinfo {author} {\bibfnamefont {S.}~\bibnamefont {Fishman}},\
  and\ \bibinfo {author} {\bibfnamefont {R.}~\bibnamefont {Prange}},\
  }\bibfield  {title} {\bibinfo {title} {Localization in an incommensurate
  potential: An exactly solvable model},\ }\href
  {https://doi.org/10.1103/PhysRevLett.49.833} {\bibfield  {journal} {\bibinfo
  {journal} {Physical Review Letters}\ }\textbf {\bibinfo {volume} {49}},\
  \bibinfo {pages} {833} (\bibinfo {year} {1982})}\BibitemShut {NoStop}%
\bibitem [{\citenamefont {Fishman}\ \emph {et~al.}(1982)\citenamefont
  {Fishman}, \citenamefont {Grempel},\ and\ \citenamefont
  {Prange}}]{fishman1982chaos}%
  \BibitemOpen
  \bibfield  {author} {\bibinfo {author} {\bibfnamefont {S.}~\bibnamefont
  {Fishman}}, \bibinfo {author} {\bibfnamefont {D.}~\bibnamefont {Grempel}},\
  and\ \bibinfo {author} {\bibfnamefont {R.}~\bibnamefont {Prange}},\
  }\bibfield  {title} {\bibinfo {title} {Chaos, quantum recurrences, and
  {Anderson} localization},\ }\href
  {https://doi.org/10.1103/PhysRevLett.49.509} {\bibfield  {journal} {\bibinfo
  {journal} {Physical Review Letters}\ }\textbf {\bibinfo {volume} {49}},\
  \bibinfo {pages} {509} (\bibinfo {year} {1982})}\BibitemShut {NoStop}%
\end{thebibliography}

%

\end{document}